\documentclass[useAMS,usenatbib]{mn2e}
\usepackage{graphicx}
\usepackage{graphics}
\usepackage{epstopdf}
\usepackage{boxedminipage}
\usepackage{float}
\usepackage{url}
\usepackage{bm}
\usepackage{epsf}
\usepackage{url}

\newcommand{\apj}{{ApJ}}
\newcommand{\apjl}{{ApJL}}
\newcommand{\apjs}{{ApJS}}
\newcommand{\mnras}{{MNRAS}}
\newcommand{\nat}{{Nature}}
\newcommand{\aap}{{A\&A}}
\newcommand{\aj}{{A\&A}}

\title[The Hierarchical Structure and Dynamics of Voids]{The Hierarchical Structure and Dynamics of Voids}

\author[Aragon-Calvo M.A. et al.]{M. A. Aragon-Calvo$^{1}$\thanks{E-mail:miguel@pha.jhu.edu} ,  Szalay A. S.$^{1}$\\
$^{1}$The Johns Hopkins University, 3400 Charles St., Baltimore, MD, USA\\}
\begin{document}

\date{Submitted to MNRAS}
\pagerange{\pageref{firstpage}--\pageref{lastpage}} \pubyear{2010}
\maketitle
\begin{abstract}

Contrary to the common view voids have very complex internal structure and dynamics. 
Here we show how the hierarchy of structures in the density field inside voids is reflected 
by a similar hierarchy of structures in the velocity field. Voids defined by dense
filaments and clusters can de described as simple expanding domains with coherent
flows everywhere except at their boundaries. At scales smaller that the
void radius the velocity field breaks into expanding sub-domains corresponding 
to sub-voids. These sub-domains break into even smaller sub-sub domains at
smaller scales resulting in a nesting hierarchy of locally expanding domains.

The ratio between the magnitude of the velocity field responsible for the expansion of the
void and the velocity field defining the sub voids is approximately one order of magnitude.
The small-scale components of the velocity field play a minor role in the shaping of
the voids but they define the local dynamics directly affecting the processes of galaxy
formation and evolution.

The super-Hubble expansion inside voids makes them cosmic magnifiers by stretching their internal primordial density
fluctuations allowing us to probe the small scales in the primordial density field. Voids also act like
time machines by ``freezing" the development of the medium-scale density fluctuations responsible for the
formation of the tenuous web of structures seen connecting proto galaxies in computer simulations. 
As a result of this freezing haloes in voids can remain
``connected" to this tenuous web until the present time.
This may have an important effect in the formation and evolution of galaxies in voids by providing an efficient 
gas accretion mechanism via coherent low-velocity streams that can keep a steady 
inflow of matter for extended periods of time.

\end{abstract}


\section{Introduction}

Voids are the most salient feature of the Cosmic Web. Early galaxy maps unveiled a far from homogeneous Universe 
dominated by vast semi-spherical empty regions 
surrounded by a complex network where galaxies aggregate into dense massive clusters joined by long filamentary bridges.
\citep{Joeveer78,Tarenghi78,Tifft78,Einasto80,Tully87,Geller89}. 
This ``Cosmic Web" can be described as an interconnected network with four basic components: 
spherical clusters connected by elongated filaments defining two-dimensional sheets forming a ``net"
that surrounds the voids that account for most of the volume in the Universe. 
This rich network was already imprinted in the primordial density field as tiny Gaussian fluctuations according
to inflationary models. 
The development of the Cosmic Web from the primordial nearly homogeneous field is a natural consequence 
of the gravitational anisotropic collapse as 
described by the seminar work of \citet{Zeldovich70}. A more explicit description of the relation 
between the different elements of the Large Scale Structure (LSS) came with the ``bubble" cosmology of \citet{Icke84} and the
``Cosmic Web" theory of \citet{Bond96}. These works among many others (see \citet{Weygaert02} for an excellent review) 
give insight on particular aspects of the anisotropic gravitational collapse that ultimately produces the
observed cosmic structures but can not fully describe the development of structure into the full non-linear regime.
In order to understand the non-linear matter evolution one usually relies on numerical experiments. N-body computer 
simulations have been used to reproduce and study the distribution of matter on large scales with great success starting 
from the pioneering work of \citep{Doroshkevich80,Mellot83,Klypin83} to the current state-of-the-art such as the millennium 
simulation with billions of mass particles \citep{Springel05b}.

%
\subsection{Studying the underdense Cosmic Web}

Voids are underdense structures that to a first approximation have simple structure and dynamics and as such their
properties are in general considered well understood. Voids are expected to have nearly spherical shapes as result of their expansion 
which erases any original sphericity \citep{Icke84}. Their internal dynamics can be well described as simple expanding domains where
$\delta \propto - \nabla \mathbf{v}$ indicating that the inner regions of the voids expand faster than the outer shells.
This gives characteristic density and velocity profiles \citep{Bardeen86} that have been observed in computer simulations 
\cite{Weygaert93, Schaap07} and observations \citep{Nasonova11}.
The work of \citet{Sheth04} represented an important step in our understanding of the dynamics of voids by
dissecting the different scenarios of void evolution based in the excursion set analysis (see also \citet{Paranjape12}).

Voids have been the subject to many studies motivated by their simple structure and dynamics, their low density and apparently
evolution \citep{Zeldovich82,Suto84,Einasto89,Kirshner81,Tully87,Szomoru93,Slezak93,Weygaert93,Szomoru96,Blumenthal92}. 
The interest in voids led to the development of many void finders (see the void comparison project \citep{Colberg08}) among which 
the watershed void finder \citep{Platen07} and the Zobov \citep{Neyrinck08a} methods stand out because of their
connection of local minima with the topopoly of the density field. 

\subsection{Galaxies in voids}

Even though voids are by definition empty in some cases one can find galaxies at their interior. In fact what is surprising is that
we do not see a large population of low-mass galaxies populating voids and tracing their inner structure \citep{Klypin99,Moore99} 
and that the galaxies we see are basically representative of the general population \citep{Peebles01}. 
Galaxies in voids have been studied in detail in the Bootes void \citep{Peimbert92,Weistro92,Weistrop95,Szomoru96b} and the
local void \citep{Karachentsev04,Wong06,Kreckel11a} and the Sloan Digital Sky Survey \citep{Patiri06,Ceccarelli08}. Recently new surveys 
targeted to voids have been carried out with primissing results \citep{Kreckel11b}. The relatively simplicity of void shapes and dynamics has been 
exploited to propose their use as cosmological probes with tantalizing results \citep{Padilla05,Lee06,Park07,Biswas10,Chongchitnan10,Lavaux10}.

\begin{figure*}
  \includegraphics[width=0.99\textwidth,angle=0.0]{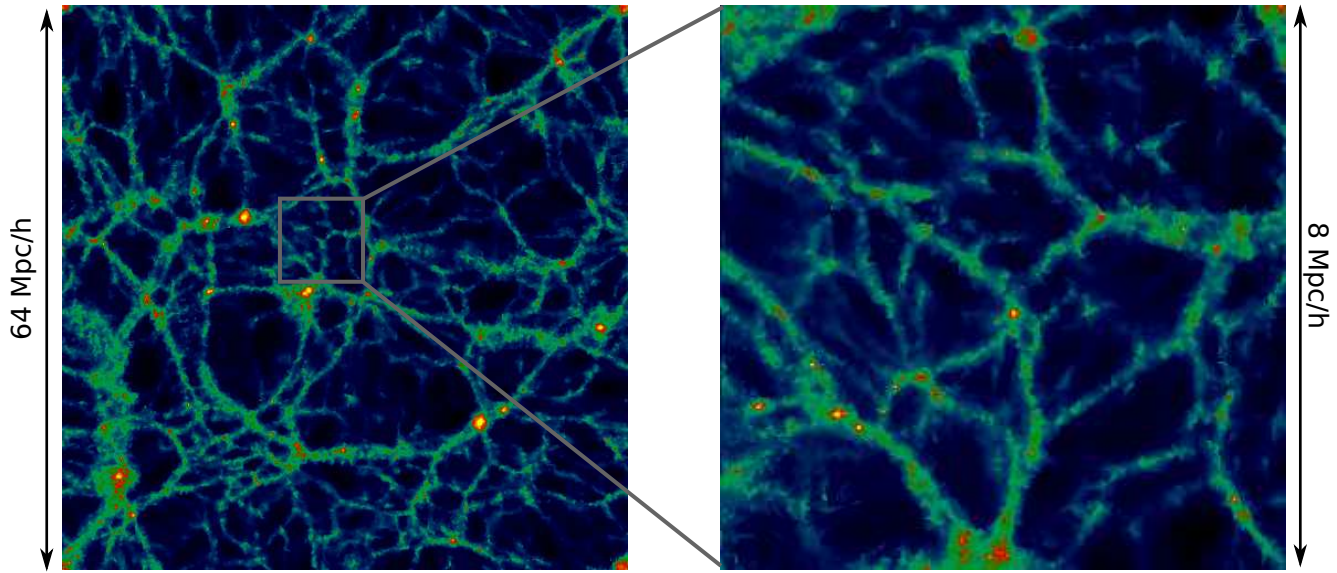} 
  \caption{Hierarchy of structure in the Cosmic Web. The left panel shows the density field on a thin slice across a 64 Mpc$h^{-1}$ simulation with
          $\Lambda$CDM concordance cosmology. The right panels show the highlighted square inside the slice. The density field in the zoomed region
		  was computed from a high-resolution resimulation with N$_{\textrm{\tiny{part}}} = 1024^3$ centered in the void region.}
    \label{fig:void_inner_web}
\end{figure*}

%
\subsection{Scope of this work}

This is the first of a series of papers exploring the hierarchical properties of the Cosmic Web. 
In the following sections we discuss the emergence and hierarchical character of the Cosmic Web as a whole.
However, in the present work we focus our analysis on voids. Walls, filaments and clusters will be treated in following papers.

%
\section{The Collapsing and Expanding Cosmic Web}

The seminal work of \citet{Zeldovich70} described the evolution of a primordial could of matter as a succession of dynamical and 
geometrical stages where the dimensionality is reduced by the gravitational collapse describe by:

\begin{equation}
\frac{\rho(x,t)}{\rho} \propto  \frac{1}{  (1- a(t)\lambda_1) \; ( 1-a(t)\lambda_2) \;( 1-a(t)\lambda_3) }
\label{eq:zeldovich}
\end{equation}

\noindent where $\lambda_3 > \lambda_2 > \lambda_1 $. Gravitational collapse occurs when at least one of the eigenvalues 
is positive. When all three eigenvalues are positive the collapse is complete, first forming flat pancakes then elongated 
filaments and finally compact dense clusters, following the decreasing order of the eigenvectors (see Table \ref{tab:zeldovich}).

\begin{table}
  \begin{center}
    \leavevmode

\begin{tabular}{|c|c|c|c|l|l|} 
\hline 
\hline 
 & $\lambda_1$ & $\lambda_2$ & $\lambda_3$ & Collapse & Expansion\\ 
\hline 
 & + & + & + & node      &  \\
 & + & + & $\ominus$  & filament & \bf{1D valley}\\
$\uparrow$ & + & $\ominus$ & $\ominus$ & wall        &  \bf{2D valley}\\
 & $\ominus$ & $\ominus$ & $\ominus$ & void         & \bf{3D valley}\\ 
\hline
\hline 
\end{tabular}

\caption{Dimensionality of a  Cosmic Web element from the negative eigenvalues of the deformation tensor. Positive eigenvalues indicate
collapse while negative eigenvalues expansion.}
  \label{tab:zeldovich}
  \end{center}
\end{table}

%
\subsection{The Expanding Cosmic Web:Inverse Zel'dovich dynamics}

The Zel'dovich equation can be also seen from the point of view of the dimensionality of the different stages in the
gravitational collapse given by the \textit{negative} eigenvalues. Voids, walls and
filaments correspond to primordial clouds with three, two and one negative eigenvalues respectively. This strongly suggest
that the dynamical character of each morphology is given by its expanding eigenvectors. In this inverse Zel'dovich picture
voids expand in all directions, walls expand restricted to their plane and filaments do similarly along their length \citep{Icke84}.
This local expansion is possible even for structures denser than the global mean such as walls and filaments 
because they are \textit{locally underdense} with respect to their surroundings filaments in the case of walls and their
surrounding dense groups and clusters in the case of filaments.
We then generalize the idea of a ``valley"  as a locally underdense $D^N$ region with voids, walls and
filaments corresponding to dimension $N=3,2,1$ respectively.
By accounting for both the positive and negative eigenvalues of the Zel'dovich collapse we have a complete picture of
a collapsing cloud (positive eigenvalues) and its internal dynamics (negative eigenvalues).

%
\subsection{Hierarchical emergence}\label{sec:hierarchical_emergence}

What happens in the Zel'dovich collapse when we consider a cloud with substructure? This is a more realistic scenario than the
idealized Zel'dovich collapse since the primordial density field is a superposition of waves.  
The hierarchical character of the Cosmic Web is imprinted in the primordial density field characterized by its power spectrum.
At the scales relevant for the development of the Cosmic Web seen in the galaxy distribution (up to a few tens of Megaparsecs) 
\citep{Einasto11} the power spectrum dictates that small fluctuations grow first superimposed
on large fluctuations naturally producing a hierarchical scenario. In the linear regime all Fourier modes are independent from
each other and grow independently (this property will be exploited in section \ref{sec:hierarchical_spaces}).

In a hierarchical scenario an over dense cloud will
collapse while internally smaller fluctuations will grow superimposed to the main collapse. This is the well known
cloud-in-cloud scenario that gives rise to haloes and and their substructure \citep{Bardeen86,Bond91,Lacey93,Sheth01}. 
On the other side of the Zel'dovich equation we have
several scenarios for the development of underdense regions as described by \citet{Sheth04} for the case of D$^3$ valleys. 
Here we have a hierarchy of voids where a given underdense region can contain locally under dense sub-regions 
or even be contained by a larger overdense region. In the first case 
the void expands while internal sub-voids also expand superposed to their parent voids's expansion. In the former case a void can 
be embedded inside larger overdense systems that eventually lead to the collapse of the void itself.

%
\subsection{The hierarchy of voids}

The distribution of matter inside voids studied with high-resolution computer simulations shows a very complex picture that is not
seen in the galaxy distribution \citep{Mathis02,Gottlober03,Colberg05a}. 
If galaxies populate the dark matter haloes found in computer simulations then one would expect to see 
the usual network of galaxies traced by large haloes and underneath it a much richer sub-web of tenuous
filaments containing a large population of low-mass haloes \citep{Dubinski93}. 

The large-scale fluctuations that give origin to cosmological voids contain also smaller fluctuations which grow inside what
behaves like a locally low-density universe. The superimposed embedded fluctuations grow faster than their parent void 
although damped by the parent void expansion. As a result of this a given D$^3$ valley will contain and be defined
by smaller D$^3$ valleys. This process occurs at all levels in the hierarchy of voids as sub-voids themselves also contain
smaller fluctuations (see Figure \ref{fig:void_inner_web}).

\subsubsection{Hierarchy of walls and filaments}
The substructure scenarios described above can be difficult to visualize in the case of walls and filaments if one only focuses on their
collapse. However, if one focuses on the negative eigenvalues then we can see that an expanding 
$D^N$ valley can contain embedded $D^N$ and $D^{N-1}$ sub-valleys which expand locally with respect to their parent expansion.
This restriction in the dimensionality of substructure comes directly from equation \ref{eq:zeldovich}
and means that a filament can have sub-filaments and nodes but no sub-walls or sub-voids. A wall can contain sub-walls
and sub-filaments but no sub-voids and a void can contain all kinds of substructure (see Table \ref{tab:zeldovich_substructure}).

\begin{table}
  \begin{center}
    \leavevmode

    \begin{tabular}{|c|c|c|c|c|} 
      \hline 
      \hline 
          & sub-voids & sub-walls & sub-filaments & \\ 

      \hline 
      void      & $\surd$ & $\surd$  & $\surd$   & D$^3$\\
      wall   &              & $\surd$  & $\surd$   &  D$^2$  \\
      filament   &            &                & $\surd$   & D$^1$ \\
      \hline 
      & D$^3$ & D$^2$ & D$^1$ \\
      \hline 
    \end{tabular}
    
\caption{Allowed substructures according to the dimensionality of the parent structure (D$^{\tiny{\textrm{N}}}$). Voids can contain all sub-elements, walls can 
contain sub-walls and sub-filaments and filaments can contain only sub-filaments.}
  \label{tab:zeldovich_substructure}
  \end{center}
\vskip 0.5truecm
\end{table}

%
\subsection{Hierarchical velocities in voids}

The velocity field in the Universe is dominated by large streaming or bulk motion. Our own Mily Way together with the local group 
is moving with a speed of $\sim 600$ km/s towards the great attractor \citep{Shaya84,Dressler87,Lynden88,Aaronson89} and at 260 km/s 
in the opposite direction from the local void \citep{Tully08}. At medium scales he dynamics around the Milky 
Way are affected by the local expansion of our parent wall, producing the so-called ``cold Hubble flow" \citep{Aragon11a}.
The power spectrum of the velocity field $P_v(k)$ is given in terms of the power spectrum of the density field $P_m(k)$ as:

\begin{equation} \label{eq:vel_power}
	P_v(k) \propto k^{-2} P_m(k).
\end{equation}

\noindent The $k^{-2}$ boost the power at large scales while the small scales have very little power in comparison.
From this we see that the velocity field is a very good indicator of the large scale matter distribution. The velocity
fluctuations at a given $k$ will be superimposed on much larger fluctuations in the form of bulk-flows which originate
from larger density fluctuations.
In fact the velocity field correlates poorly with the observed features in the cosmic web 
(see top panels in Figure \ref{fig:hierarchical_velocities}). 
The $k^{-2}$ factor effectively produces a mismatch between the dominant scales in the density field and the velocity field.

Consider a filter that applied to the velocity field will attenuate the power at large scales such as
$\mathcal{W} \propto k^2$ in which case $P_v(k) \propto P_m(k)$. Now the fluctuations in the velocity field 
have the same scaling as the fluctuations
in the density field and this will be reflected in a closer match between the two. Another option is a Gaussian
high-pass filter. This filter attenuates the amplitude of $P_v(k)$ at large scales bringing the shape of the power spectrum closer 
to $P_m(k)$ after the cut-off scale. A high-pass filter provides a way to select scales in the velocity 
field associated to particular elements in the Cosmic Web. 
The high-pass filtered velocity field contains only local fluctuations at the cut-off scale. This is analogous to measuring peculiar velocities after
removing the averaged local velocity around a window of a given size. Since the high-pass filter matches the velocity and mass power spectrum
we will be able to identify small scale features in the density field reflected in the velocity field. 
We should then expect to see a hierarchy of expansion domains at the interior of voids reflecting the
hierarchy of voids observed in the matter distribution.

Equation \ref{eq:vel_power} is valid in the linear regime. At the present time small-range gravitational interactions make the
peculiar velocities in the non-linear, high-density environments very high and turbulent
as power is transfered from large scales.
On the other hand, the low-density nature of cosmological voids  results in a
velocity field dominated by large laminar motions and at smaller scales low-velocity coherent streams. This
makes voids ideal laboratories to study the velocity field at all scales.

\subsubsection{Residual velocities}\label{sec:residual_velocities}

In order to isolate the local component of the velocity field from the bulk flows that dominate the overall
dynamics of the Cosmic Web we define the \textit{residual velocity} at scale $r$ as:

\begin{equation}
   \mathbf{v}_{res}(r) = \mathbf{v} - \mathbf{v}_{G(r)}
\end{equation}

\noindent where $\mathbf{v}$ is the original velocity field and $\mathbf{v}_{G(r)}$ is the velocity field smoothed with a Gaussian filter 
of radius $r$. This is a high-pass filter that suppresses flows at scales larger than $r$ and allows us to probe the velocity field at an 
specific local scale. In the following sections we will use this filter to explicitly expose the hierarchy of velocities in voids.

\subsection{Voids as cosmic microscopes and time machines}

Voids offer a unique opportunity to probe the small scales in the primordial power spectrum and the early development of
medium-scale fluctuations. The small-scale density fluctuations embedded in voids are stretched as the void expands, 
effectively turning voids into cosmic microscopes. Inside these low-density environments the growth of primordial fluctuations is
attenuated and continue in the linear and semi-linear regimes
for longer time than in denser regions. Voids then offer a direct way to study the primordial
power spectrum at smaller scales than in their complementary high density regions where the opposite effect occurs
and small fluctuations are compressed and even erased by the gravitational collapse.

There is a perhaps more interesting property of the substructure inside voids originating from the
break in symmetry in the collapse of small fluctuations depending on their position with respect to
larger fluctuations. Following along the same lines as  \citet{Suhhonenko11,Einasto11,Einasto11b} lets consider the evolution 
of a given set of small-scale fluctuations in different large-scale environments.
If the small-scale fluctuations are located on top of a large overdense peak they will begin to grow enhanced by their position 
inside the peak and at the same time defining a full medium-scale cosmic web connecting the small-scale peaks
(see figure 3 in \citet{Romano11}). 
The larger volume of the peak will eventually collapse erasing any memory of the primordial 
fluctuations and their medium-scale web. On the other hand, if the same small-scale 
fluctuations are located on a large underdense valley their growth will be attenuated by the expansion of the void
but more important, the expansion will stretch the fluctuations and they will never collapse as a whole as in the case of the large peak. 
Instead they will evolve into a tenuous web with its own network of voids, walls, filaments and low-mass haloes. 
The expansion of the void will further stretch this inner web to larger scales (see Figure \ref{fig:void_inner_web}). 
The end result is that the band of $k$ in the power spectrum responsible of producing 
the tenuous cosmic web inside voids is effectively shifted towards higher values of $k$. 
Voids then offer an unique opportunity to study the early evolution of the ``frozen" cosmic web at medium-scales. 
The same idea can be applied to walls and perhaps even to less extent to filaments, although the fluctuations there correspond to 
the power spectrum collapsed along one and two directions respectively, making the interpretation not as direct as 
in the case of voids.

\section{Deconstructing the Universe: The Cosmic Spine}

The Hierarchical Cosmic Spine is the most recent implementation of the original method introduced in
\citet{Aragon10a} and \citet{Aragon10b}. The main improvements presented here are the addition of explicit hierarchical 
relations for voids, walls and filaments and a self-consistent criteria for the identification of cosmic structures.

We start by identifying the voids in the density field using a floating-point implementation of the original watershed void finder
\citep{Platen07}. This provide us with a partition of space into contiguous regions sharing the same local minima. In the 
Cosmic Spine method \citep{Aragon10a} we assign the boundaries between the watershed regions 
(i.e. the watershed transform) to the network of walls, filaments and clusters.

From the topology-based segmentation provided by the Watershed Void Finder we characterize the local geometry of
the density field as follows: We start by labeling regions identified with the watershed transform as
cosmological voids. Each void has its own label provided by the watershed transform. We then focus on the boundaries
between void regions and label the intersection between two voids as walls. Then the intersection between walls correspond
to filaments and the intersection between filaments are the nodes of the network which in the Universe correspond to
galaxies, groups and clusters (see Table \ref{tab:spine_criteria}). This is done in practice by iterating 
through all the voxels labeled as watershed boundaries and then
labeling as walls those voxels that have two different voids in their adjacent $3 \times 3 \times 3$ neighborhood. We run again
over the remaining unlabeled boundary pixels and identify filaments as voxels have at least two walls in their adjacent neighborhood.
Finally, we identity haloes as voxels having two or more adjacent filaments. 
Even though our method provide us with haloes as the 
nodes of the network we opt to identify them using other methods like FoF that are not limited by the grid resolution. 
The Cosmic Spine method provides with a complete framework for the characterization of the cosmic web into its 
primary constituents: voids, walls, filaments and clusters. It has no free parameters (with the possible exception of grid
resolution which is not an intrinsic free parameter since its variation does not affect the final result as long
as the spacial sampling can recover the features in the density field) 
and being fundamentally a topological measure it is highly robust agains noise and artifacts in the density field. 

\begin{table}
  \begin{center}
    \leavevmode

\begin{tabular}{c c l | l} 
\hline
\hline
\multicolumn{3}{|c|}{Adjacency condition}     & Structure  \\ 
\hline 
           & 2 & voids        &    wall \\
 $\ge$ & 2 & walls        &    filament  \\
 $\ge$ & 2 & filaments  &    node \\
\hline 
\hline 
\end{tabular}
\caption{Criteria used to classify structures in the Cosmic Spine method.}
  \label{tab:spine_criteria}
  \end{center}
\vskip 0.5truecm
\end{table}

\subsubsection{Hierarchical SpineWeb}\label{sec:hierarchical_spine}

The main limitation of the original Cosmic Spine method is the lack of control in the structures that we identify. A phenomenon
referred to as \textit{oversegmentation}. Being a topological method
any feature in the density field that affects its connectivity will be detected regardless its contrast. 
For instance, consider the 2D case of a void with very dense boundaries but an extremely 
tenuous bridge of matter across its center. The Spine method will identify two voids as 
long as there is a bridge of matter dividing the slope lines emanating from the
local minima. Several approaches have been proposed to deal with these substructures inside voids 
all involving the use of a free parameter as some form of threshold or significance measure \citep{Platen07, Sousbie11}. 
We take different approach and first consider
the nature of the substructure. In N-body computer simulations where we have complete control over the sampling of the
underlying density field (particles in this case) we expect that the fidelity of the features in the density field is limited by the 
accuracy of the method we use to estimate the density and that (beyond numerical accuracy) any feature in the density field 
is the direct result of gravity. From the discussion in Section \ref{sec:hierarchical_emergence} we see that substructure is a natural consequence of the hierarchical 
development of structure and therefore we should not remove it by using ah-hoc procedures. Instead the proper
way to deal with such hierarchy of structures is by using a hierarchical approach to the identification of structures.
This is the main motivation behind the Hierarchical Spine Web. By doing a hierarchical classification we avoid the need
of an extra free parameter and instead consider the Cosmic Web as a network of nesting structures where
large structures contain and are defined by smaller ones.

\begin{figure*}
  \includegraphics[width=0.99\textwidth,angle=0.0]{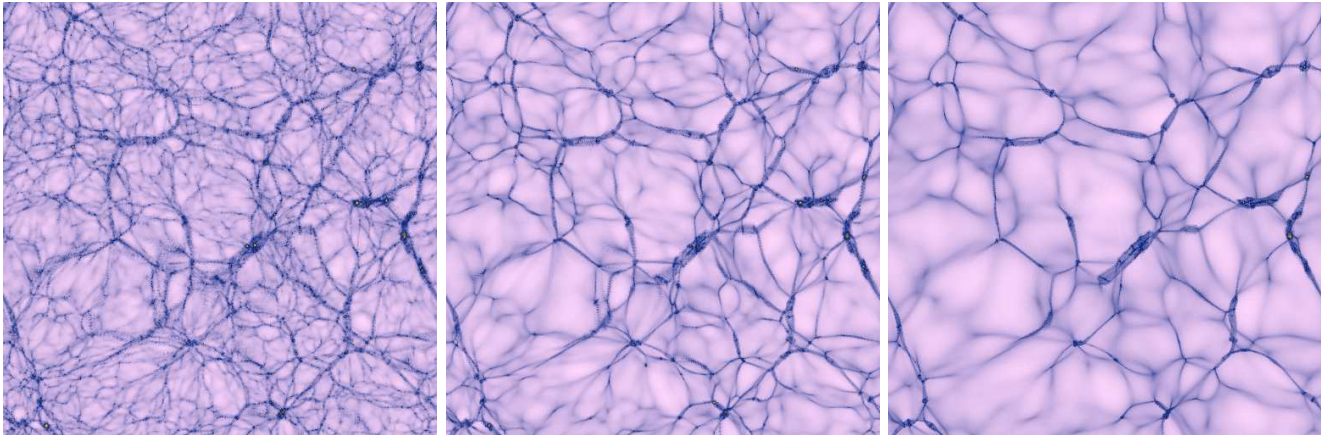} 
  \caption{Density field across a thin slice of the simulation box at $z=0$ after linear-regime smoothing for three cases:
  original field (left panel), 2 $h^{-1}$Mpc (central panel) and 4 $h^{-1}$Mpc (right panel).}
    \label{fig:multires_density}
\end{figure*} 

%
\subsection{Constructing hierarchical Spaces} \label{sec:hierarchical_spaces}

The crucial point in the hierarchical analysis is the selection of a proper filter that can expose the hierarchy of structures in the Cosmic Web.
This is accomplished by doing the Cosmic Spine analysis in hierarchical-space \citep{Aragon10b} which is a generalization 
of scale-spaces that are defined by one of more properties that are manifestations of the hierarchical nature of the Cosmic Web.
Hierarchical spaces are not restricted to linear scale-space operators such as the Gaussian kernel
\citep{Florack93}.
In the case of N-body simulations we can take advantage of having access to the full evolution of the simulation to design a
filter that targets specific scales in the linear regime at early times when the Fourier modes of the density field are independent 
and grow independently (see Section \ref{sec:hierarchical_emergence}). This linear-regime low-pass filtered density field will 
evolve into a Universe with all the structures
above the cut-off scale in place and shaped by the anisotropic gravitational collapse, but lacking
of the small-scale details. This approach is fundamentally different from the usual a posteriori smoothing operation, in that 
it avoids the nonlinear effects resulting from cross talk between Fourier modes. It has the 
advantage of cleanly exposing the hierarchy of structures imprinted in the initial density field.
The linear-regime filter has a different impact on each element of the Cosmic Web. In the case of voids the
initial fluctuations will be smoothed at a given comoving scale in the linear regime but as the void evolves and expands
the embedded fluctuations will be stretched, effectively increasing the scale of the smoothing length. In the case of clusters
we have the opposite effect with the matter inside the turnaround radius collapsing and erasing the small scale fluctuations.
Filaments and walls are intermediate cases.

Figure \ref{fig:multires_density} shows the density field at $z=0$ evolved from the original initial conditions (left panel) and 
linear-regime smoothed versions at  1 $h^{-1}$Mpc  and 2 $h^{-1}$Mpc (center and right panels respectively). The 
linear-regime smoothed versions lack of small scale details. Even though there are no small scale features the large voids still 
have well defined sharp boundaries in contrast to what we would see in a Gaussian smoothing. At the top
of the hierarchy we only have large scale structures and basically featureless voids. The middle level shows some features appearing
inside the voids while the bottom level shows a rich inner-web at the interior of the voids. This progressive addition of features is 
exploited in the following sections to identify and characterize the full hierarchy of structures in the cosmic web.

\begin{figure*}
  \includegraphics[width=0.99\textwidth,angle=0.0]{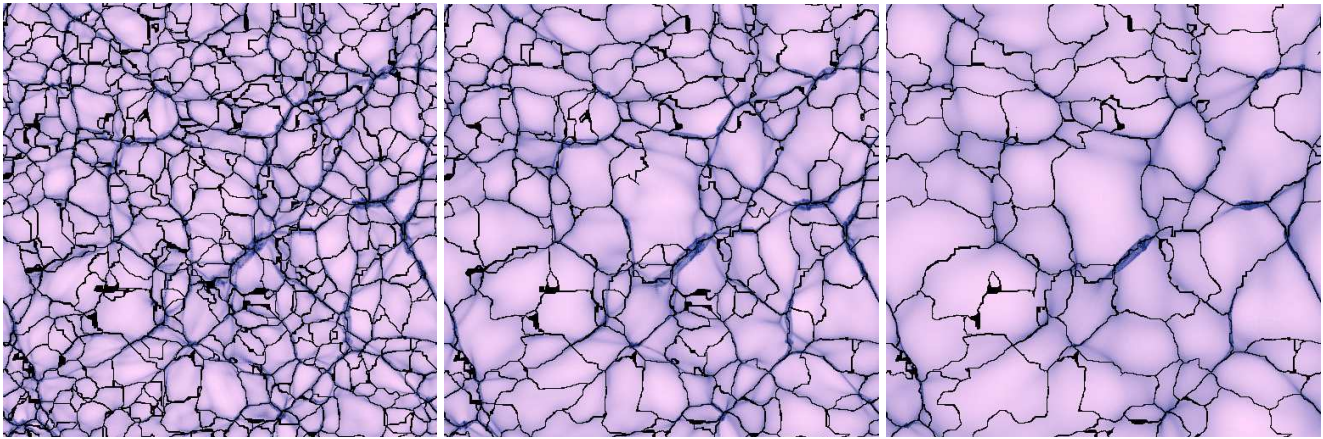} 
  \caption{Hierarchical Cosmic Spine superimposed to its corresponding density field for the bottom, middle and top 
  levels (left, central and right panels respectively). The black lines delineate the boundaries of the voids found
  with the Hierarchical Cosmic Spine. The background corresponds to the 
  density field across a thin slice of the simulation box at $z=0$ after linear-regime smoothing for three cases:
  original field (left panel), 2 $h^{-1}$Mpc (central panel) and 4 $h^{-1}$Mpc  (right panel).}
  \label{fig:multires_spine}
\end{figure*} 

%
\subsection{Hierarchical void finding} \label{sec:hierarchical_void_finding}

We start the hierarchical characterization of the Cosmic Web by identifying voids regions as described in the previous section.
In the Cosmic Spine hierarchical framework we identify voids independently at all levels in hierarchical space, 
and establish the cross-scale relations between voids at different levels. 
A given {\it parent} void at the hierarchy level $i$ is defined by smaller {\it children} voids at the next hierarchical level, $i+1$. 
We assign parent-child relations by identifying overlapping volumes between voids in adjacent levels in the hierarchical space.
A given child usually shares volume with several parent voids higher in 
the hierarchy. We enforce a {\it non-loop} property in the hierarchical tree by assigning each child void exclusively 
to the one parent void to whom the child contributes most of its volume. This constraint assures that all children 
voids have only one single parent in the \textit{void tree hierarchy}. 

We perform the void merging across adjacent levels in the hierarchy by computing only the flooding procedure on the watershed method
(see \citet{Aragon10a} for details).
This procedure segments the density field into watershed basins but 
does not explicitly provide the boundaries between adjacent watershed regions. The ``incomplete watershed" focuses 
on the space partitioning aspect of the watershed transform. This makes it straightforward to merge voxels between 
children and parent voids producing complete hierarchical void tree. 
The parent void is then reconstructed as the union of its children voids as:

\begin{equation}
\label{eq:void_union}
\mathcal{V}_p = \bigcup v_i
\end{equation}

\noindent where $\mathcal{V}_p$ is the set of voxels defining a parent void and $v_i$ are the voxels defining each void child 
contained by  $\mathcal{V}_p$. Note that the basins of the incomplete watershed also contain the watershed transform.
After the merging procedure we compute the full watershed transform 
(i.e. watershed basins and boundaries) using the reconstructed parent void basins given by eq. \ref{eq:void_union} and 
the density field with the highest resolution, corresponding to the bottom level of the hierarchy. This new watershed 
contains only large structures corresponding to the parent voids while conserving the small scale features present in the
high resolution density field. The particular watershed implementation we use  is based on a local flooding restricted to the voxels 
adjacent to void boundaries. This reduces the number of voxels needed to sort in the flooding step and greatly improves performance.
We can fully analyze a set of three $512^3$ voxel grids in a few minutes using a standard workstation.

%
\subsection{Hierarchical Cosmic Spine}

From the hierarchy of voids described in Section \ref{sec:hierarchical_void_finding} we proceed to identify walls, and filaments 
using the criteria outlined in table \ref{tab:spine_criteria} for \textit{each} hierarchical level independently. 
The hierarchical relationship between elements of the Cosmic Web in adjacent hierarchical levels
is implicitly given by the reconstruction of parent voids from their children voids so there is no need to explicitly 
establish parent-child relations again for walls and filaments. 

This hierarchical identification of structures results in a watertight network of voids surrounded by walls and filaments. 
Each void contains an inner network of sub-voids each with its own surrounding system of sub-walls and sub-filaments. 
A more detailed description of this procedure will be given in the following papers of this series. In the present work 
we focus only on the identification of voids.

\begin{table}
  \begin{center}
    \leavevmode

\begin{tabular}{l c l} 
\hline
\hline
Name & N$_{\tiny{\textrm{part}}}$ & Effective scale \\
\hline 
100Mpc-A0  & $512^3$  &  0.25 Mpc$/h$ (Nyquist)\\
100Mpc-A1  & $512^3$  &  1 Mpc$/h$ \\
100Mpc-A2  & $512^3$  &  2 Mpc$/h$ \\
100Mpc-A4  & $512^3$  &  4 Mpc$/h$ \\
100Mpc-B0  & $256^3$  &  0.4 Mpc$/h$ (Nyquist)\\
100Mpc-C0  & $128^3$  &  0.4 Mpc$/h$ (Nyquist)\\
100Mpc-D0  & $64^3$    &  0.8 Mpc$/h$ (Nyquist)\\
\hline 
\hline
  \end{tabular} 
\caption{Effective scales of the simulations used in this work. The effective scale indicates the cut-off scale
given by the linear regime Gaussian smoothing or the natural scale given by the particle sampling.
The suffix 0 in the name indicates initial conditions where the particle sampling corresponds to the Nyquist limit.}
  \label{tab:simulations}
  \end{center}
\vskip 0.5truecm
\end{table}

\begin{figure} 
  \includegraphics[width=0.49\textwidth,angle=0.0]{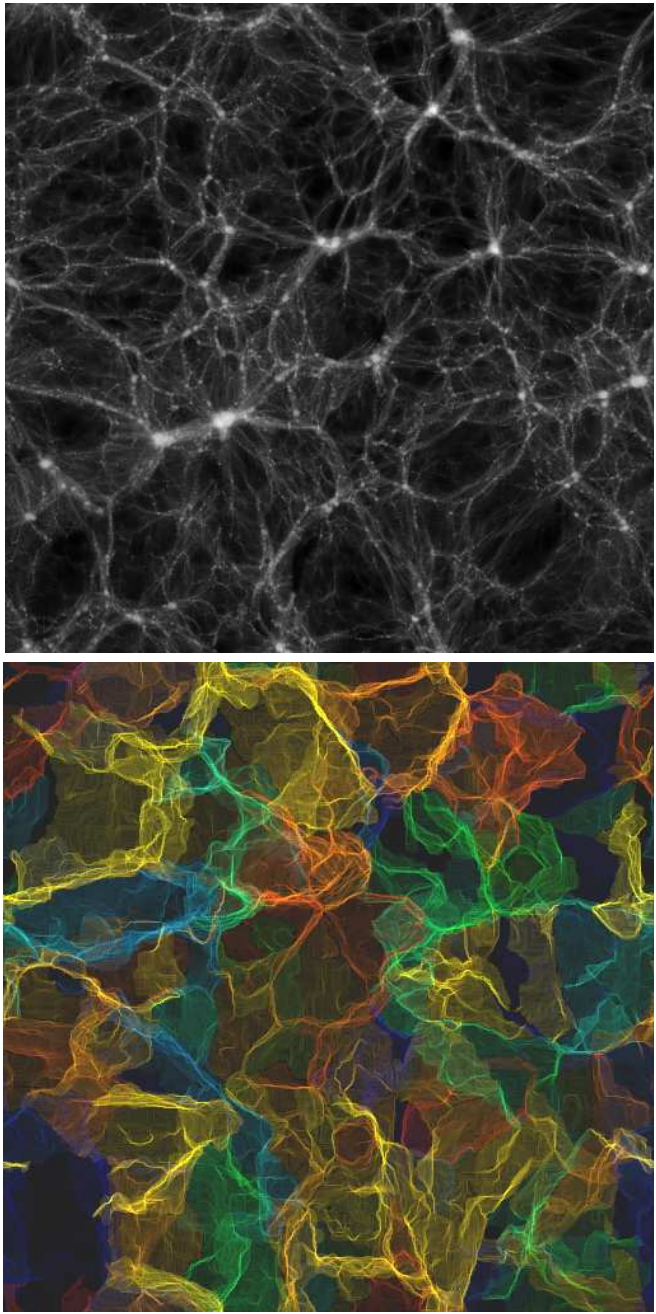} 
  \caption{Density field across a thin slice of the simulation box (top panel) and its corresponding void network (bottom panel). 
		Only the boundaries of the voids are shown as semi-transparent membranes each with a different randomly assigned color.
		We show the full tridimensional extent of all voids that are intersected by the density slice on the top panel.
		The boundaries were visualized by volume rendering the pixels in the boundaries of the voids with a random color
		transfer function.}
  \label{fig:void_network}
\end{figure}

\section{Simulations}

The analysis presented here is based on an N-body simulation with dark matter corresponding to the
standard cosmology with cosmological parameters $\Omega_m = 0.3, \Omega_{\Lambda} = 0.7, 
h = 0.73, \sigma_8 = 0.8$. Initial conditions were generated for $512^3$ particles in a periodic box of
100 Mpc$h^{-1}$ using the publicly available \verb+cosmics+ software \citep{Bertschinger95} with power spectrum
computed from an analytical fit.
For practical purposes like fast visualization and preeliminar analysis we generated lower resolution versions 
at  $256^3$, $128^3$ and $64^3$ particles following the averaging procedure described in \citet{Klypin01}.

%
\subsection{Linear-regime smoothing}

In order to compute the hierarchical Cosmic Spine we generated linear-regime smoothed versions
of the initial conditions as described in Section \ref{sec:hierarchical_spaces}. We take a different approach to the one presented
in \citet{Aragon10a} were we kept the particle sampling at the Nyquist frequency by merging adjacent
particles and effectively lowering the particle count by a factor of 2 on each dimension at each smoothing step.
In the present analysis we perform the linear-regime smoothing on the original initial conditions using a Gaussian window
while keeping the particle number constant. 
In practice we assign positions to particles on a regular grid using the smoothed displacement field as given in the 
Zel'dovich approximation \citep{Zeldovich70}:

\begin{equation}
	\mathbf{x} = \mathbf{q} + \mathbf{\Psi}_{\sigma}(\mathbf{q},z).
\end{equation}

\noindent  Where $ \mathbf{\Psi}_{\sigma}(\mathbf{q},z)$ is the perturbation field smoothed with a
Gaussian function at scale $\sigma$. Velocities are assigned according to:

\begin{equation}
	\mathbf{v} = f(z) \; \mathcal{H}(z) \;  \mathbf{\Psi}_{\sigma}(\mathbf{q},z).
\end{equation}

\noindent Where $f(z)$ is the logarithmic derivative of the growth factor and $\mathcal{H}$ is the reduced Hubble parameter.
With the above prescription we generated linear-regime smoothed initial conditions at 1, 2 and 4 $h^{-1}$Mpc 
(see Table \ref{tab:simulations}).
We then followed the evolution of each of the initial condition boxes from $z=49$ until the present  time, $z=0$, 
using the GADGET-2 N-body code  \citep{Springel05} (see Figure \ref{fig:multires_density}). 

%
\subsection{Computing Density field}

From the final particle distribution we compute the density field for the three simulations described in the previous section. 
This is done usind the Delaunay Tessellation Field Interpolator (DTFE) method \citep{Schaap00, Schaap07, Weyschaap09}.
The DTFE method is a parameter-free density estimator and interpolator. It assigns densities to particles
acoording to the inverse of their adjacent Voronoi cell as:

\begin{equation}
	\rho(\mathbf{x}) = \frac{(\textrm{D}+1) \; m}{ V(\mathcal{W}_i)}
\end{equation}

\noindent where $V(\mathcal{W}_i)$ is the volume of the adjacent Voronoi cell and $m$ is the mass of the particle. 
The ability of the DTFE method to follow the intrinsic anisotropies in the density field is crucial for the correct
identification of structures by the Cosmic Spine.

%
\subsection{Computing Velocity fields}

We interpolate the velocity field into a regular grid by means of the Delaunay tessellation  
\citep{Bernardeau96, Schaap07, Romano07}. The velocities at each particle in the simulation are assigned to 
its corresponding vertex in the Delaunay tessellation of the particle distribution. The velocities are then 
interpolated from the $D+1$ Delaunay vertices (per tetrahedron) $\mathbf{r}_0,\mathbf{r}_1, \mathbf{r}_2 ... \mathbf{r}_N $
at the central position $\mathbf{r}$ of every pixel on a regular grid as:
\begin{equation}
v(\mathbf{r}) = v(\mathbf{r}_0) + \widehat{\nabla v} \vert_{\tiny{\textrm{Del}}} \cdot (\mathbf{r} - \mathbf{r}_0)
\end{equation}
\noindent where $\widehat{\nabla v} \vert_{\tiny{\textrm{Del}}}$ is the field gradient inside the Delaunay tetrahedron.

In order to avoid aliasing from multistreaming when sampling velocities on a regular grid we smooth the velocity field 
per particle at the corresponding voxel size (Nyquist limit) by performing a weighted mean as:

\begin{equation}
\bar{v} =  \frac{ \sum^n_i w_i v_i} { \sum^n_i w_i } 
\end{equation}

\noindent where the weights $w_i$ are computed from a Gaussian function:

\begin{equation}
w_i = \frac{1}{ \sqrt{2 \pi \sigma^2}} e^{ \frac{ r_i^2 }{ 2 \sigma^2} }
\end{equation}

\noindent and $r_i$ is the distance to particle $i$. The weighted averaging is done separately per velocity component.
This smoothed velocity field is then used to interpolate on a regular grid using the Delaunay tessellation
as described above.

\begin{figure*}
  \centering
  \includegraphics[width=0.99\textwidth,angle=0.0]{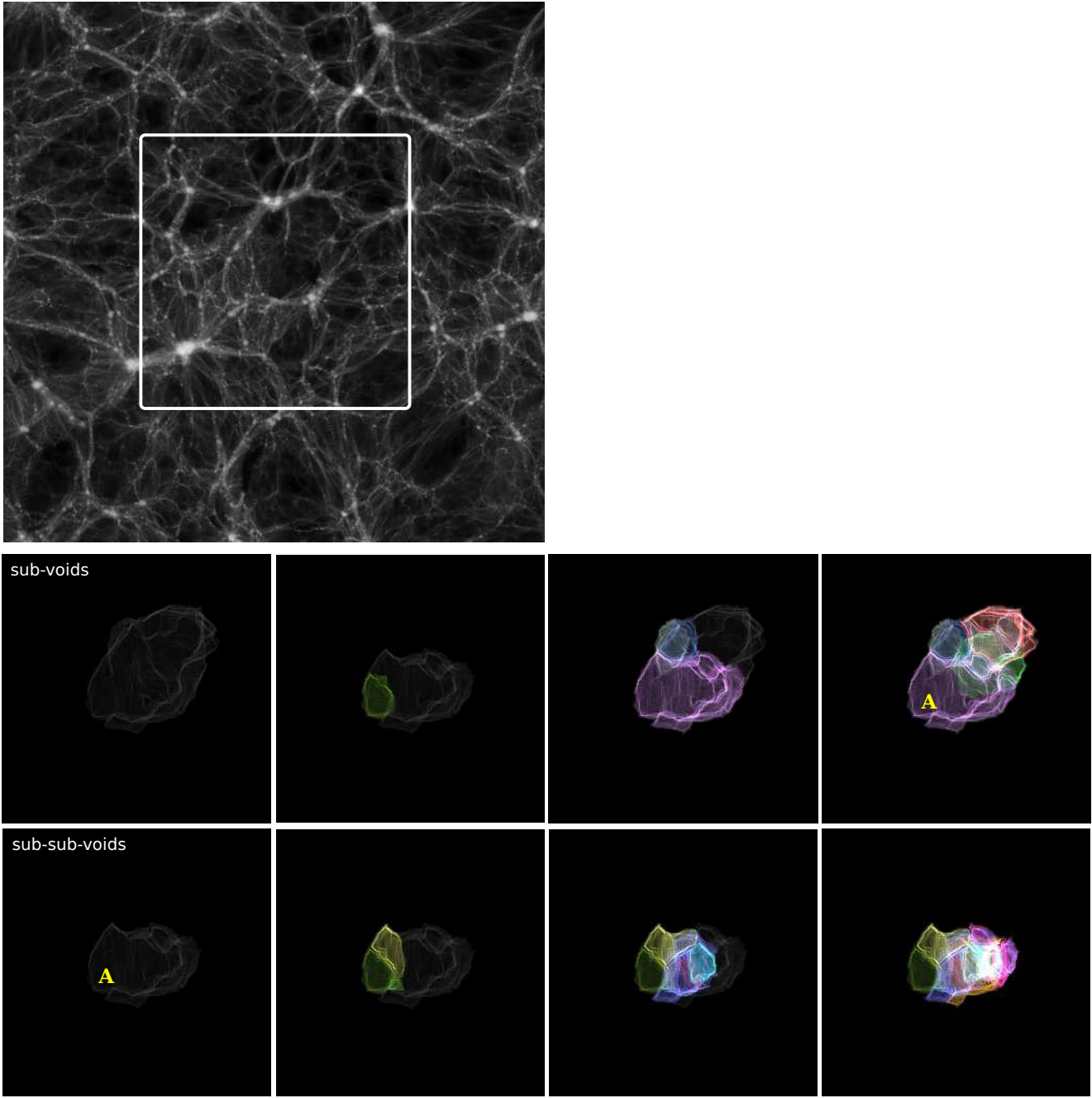}  
  \caption{Hierarchy of sub-voids inside a parent void. The top panel shows a slice across the density field from which
  		the central region with void is highlighted with a write square. The central void is then decomposed into its embedded
		sub-voids in the middle panels. The tenuous gray surface corresponds to the boundaries of the parent void
		while the colored surfaces to the sub-voids. We show three sets if sub-voids with the right middle panel
		showing all the sub-voids contained in the parent void. Note that the parent void contains and is defined by its
		sub-voids. Bottom panels focus on the largest sub-void (marked with a letter \textbf{A} in the middle right panel) and decomposed 
		it on its sub-sub-voids in a similar way as in the middle panels.}
\label{fig:hierarchy_sub_voids}
\end{figure*} 

\section{The Hierarchy of voids}

We perform the hierarchical Cosmic Spine decomposition using the initial conditions with 0.2 $h^{-1}$Mpc (Nyquist) 
and with linear-regime smoothing at 2 and 4 $h^{-1}$Mpc corresponding to simulations 100Mpc-A0, 100Mpc-A1 and 100Mpc-A2
(see table \ref{tab:simulations}). Ideally one would want to create a continuous hierarchical space. In practice, however
this is not possible and one must decide the number and spacing of hierarchical levels beforehand. 
The design of scale/hierarchical spaces is a non-trivial task and
in general the scale-space resolution is limited by the available computational resources 
(see \citet{Sato98} for an extended discussion). In our case the creation of the hierarchical space involves the
full evolution of a $512^3$ particles simulation which limits the number of levels that one can produce and analyze in a
reasonable amount of time. In our experience three hierarchical levels give a good compromise between scale coverage and the computational
resources needed. Note that in the case of voids the dynamic range
of the attenuated scales covered by the hierarchical space is larger than the initial range imposed by the
linear-regime smoothing because the scale of the density fluctuations is amplified inside 
voids giving a larger effective smoothing length. 

In what follows we denote the hierarchical levels obtained from the original density field and the linear-regime smoothed field
at 2 and 4 $h^{-1}$Mpc as top, middle and bottom levels respectively.

Figure \ref{fig:void_network} shows the void network identified at the top of the hierarchy. The top panel shows a slice through the
density field and the bottom panel shows the network of voids that cross the slice. For clarity we show each void with a different color.
The void network covers all the space in the Cosmic Web. The net of clusters, filaments and walls lies at 
the boundaries of the void network. Voids have complex shapes, not only departing from the ideal sphericity but also their boundaries
have a very complex ``texture" reflecting the non-linear nature of the dense boundaries. 
It may be surprising that voids are not regular spheres as one would expect for an expanding region \citep{Icke84}.
However, this is valid only for an \textit{isolated} void. In the context of the Cosmic Web the shape of an expanding
void is affected by the expansion of its adjacent voids and even more complex processes like 
the collapse of the void itself (the void-in-cloud scenario) in which case the shape of the void would be highly elongated \citep{Sheth04}.

The hierarchical decomposition of a single void is illustrated in Figure \ref{fig:hierarchy_sub_voids}.  The top panel shows the density field
across a thin slice of the simulation box. A white box highlights an interest region with a void at its center. The central void
has well defined dense boundaries and two large clusters almost at opposite directions. 
This particular void has significant substructure across the slice. 
In order to illustrate the hierarchical decomposition of voids we select the central void in Figure \ref{fig:hierarchy_sub_voids} (top panel) and show the
sub-voids it contains by successively adding them to their parent void (middle panels). The boundaries of the parent void are shown
as a thin gray surface and the sub-voids are shown as colored surfaces highlighting their boundaries. We then take
the largest sub-void (marked as \textbf{A}) and decompose it in its sub-sub-voids (bottom panels). Here the sub-void acts as a parent 
void in this hierarchical level and is shown as a thin gray surface. We add sub-sub-voids as colored surfaces just as in the medium panels.
There is a large variation in the relative size of the sub-void with respect to their parent void. The selected void has
three large sub-voids (purple, green and red) and several smaller sub-voids. The relative size of sub-sub-voids seems to be 
more regular. 

%
\subsection{Void size distribution}\label{sec:void_sizes}

\begin{figure} 
  \includegraphics[width=0.49\textwidth,angle=0.0]{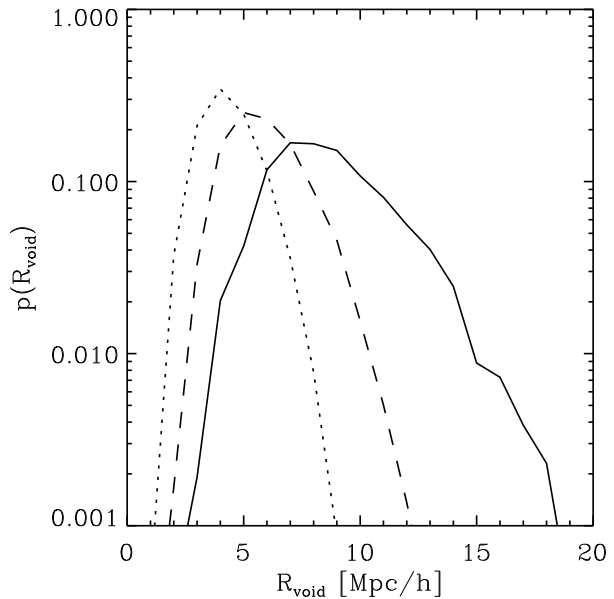} 
  \caption{Probability distribution function of void radius for the top, middle and bottom hierarchical levels computed in this
  	work corresponding to solid, dashed and dotted lines.}
  \label{fig:void_sizes}
\end{figure}

The distribution of void sizes is shown in Figure \ref{fig:void_sizes}. We estimate the void radius from its total volume as 
R$_{void} =  ( 3 / (4 \pi) V )^{1/3} $. Where $V$ is the volume of all the voxels in the void. The distribution of void sizes is shown for the three hierarchical 
levels computed in this work. The three curves have similar shape and are close to a log-normal distribution. 
The peak of the distribution increases as we move across the void hierarchy at $\sim$ 4, 6 and 8 $h^{-1}$Mpc for bottom, 
middle and top levels respectively. The peak of the top level voids is $\simeq 11$ Mpc for $h=0.73$. Recently estimates
of the radius of the local void by \citet{Nasonova11} give R$_{\textrm{\tiny{LV}}} \simeq$ 10 Mpc. According to our results
the local void is a fairly common void in terms of its size.
The largest voids we identify in our simulation correspond to the 
top of the hierarchy and have a  maximum radius close to 20 $h^{-1}$Mpc. The medium and bottom levels have maximum cut-off 
radius of 12 and 9 $h^{-1}$Mpc respectively while we can find very small voids in all levels of the hierarchy. 
The top level voids in Figure \ref{fig:void_sizes} are equivalent to the voids identified by \citet{Platen07} and in fact our results 
are in excellent agreement with their single-level analysis.
The top of the hierarchy has a very strong correlation with the voids delineated by the dense filaments and clusters of galaxies.
Combining our results with the excursion set formalism derived by \citet{Furlanetto06} we expect the voids in the three hierarchical
levels to be delineated by galaxies with a range in luminosity of M$_r < -16$ and M$_r < -20$ between the bottom and top levels respectively.

\begin{figure}
  \centering
  \includegraphics[width=0.49\textwidth,angle=0.0]{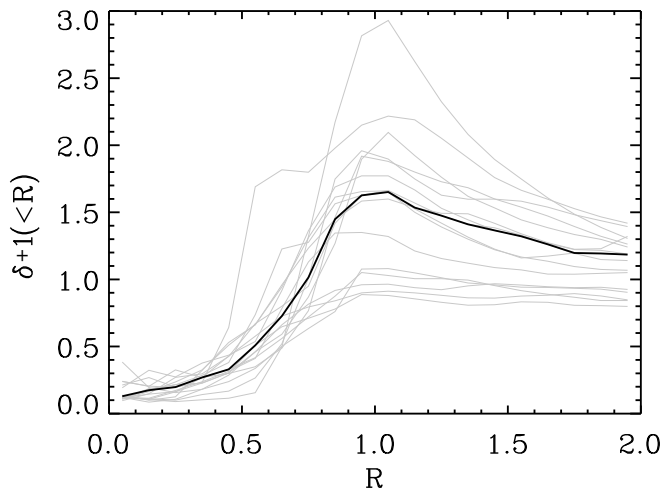}
  \caption{Density profile of voids as function of enclosing radius. The light-gray lines show several individual void profiles
  	while the black line corresponds to the mean of all the profiles. The $x$-axis of the individual profiles are normalized 
    with the peak of the curve..}
\label{fig:void_profile}	
\end{figure}

%
\subsection{Density profile of voids}

Figure \ref{fig:void_profile} shows a first impression of the internal structure of voids. Here the density profile was computed
inside increasing spheres placed at the center of voids. The void centers were computed as the mean of all the
voxels defining the void. The $x$-axis is scaled such that the density profile for each individual void reaches its peak at R=1
for easy comparison between different voids. This also allows us to stack several void (gray lines) and compute
its mean profile (black line). The density profile starts around $\delta+1 \sim 0.1$ at the center of the voids and then increases 
until it reaches the enclosing network of walls, filaments and clusters at the peak of the density profile. After this point the profile 
slowly converges to the mean density. Note that some profiles have their peak close to the mean density and some even don't seem
not reach it. These are shallow voids that are embedded in very large underdensities of the order of $\sim$ 100 $h^{-1}$Mpc and 
structures that are still in the process of formation \citep{Einasto97,Sheth04}. The density profile inside voids is the result 
of the draining of matter from the void into its dense boundaries as they collapse and increase their density high above the mean.
The void profiles we computed are in good agreement with theoretical expectations  
\citet{Hoffman83, Hausman83, Fillmore84, Bertschinger85, Sheth04} and the density profile recently computed from the 
distribution of galaxies in the local void by \citet{Nasonova11}.

\begin{figure*}
  \centering
  \includegraphics[width=0.9\textwidth,angle=0.0]{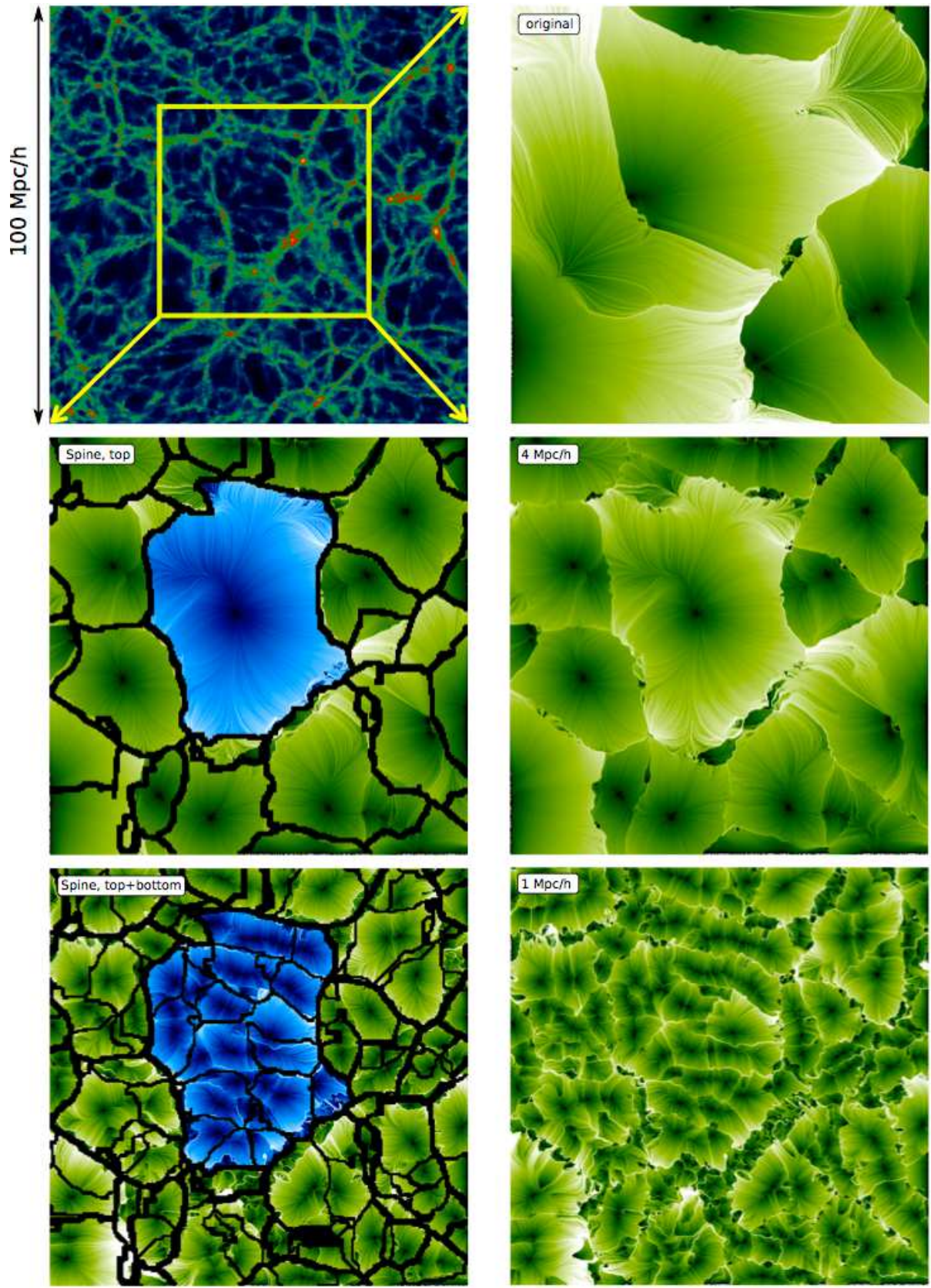}  
  \caption{Hierarchical velocities inside voids. The top left panel shows a slice of the density field from which
  the central area is selected (yellow square). 
  The right panels show the structures in the velocity field using the
  particle advection technique (see text for details) for three cases: original velocity field (right top) and 
  residual velocity field at scales 4 and 2 $h^{-1}$Mpc (middle top and right bottom panels respectively). Left panels show again
  the velocity field for 4 and 2 $h^{-1}$Mpc and superimposed the Cosmic Spine showing the top level as a thick black contour
  (middle left panel) and the top$+$bottom levels as thick and thin black contours respectively (bottom left panel). 
  For clarity we highlight the center void in blue.}
  \label{fig:hierarchical_velocities}
\end{figure*}

\begin{figure*}
  \centering
  \includegraphics[width=0.99\textwidth,angle=0.0]{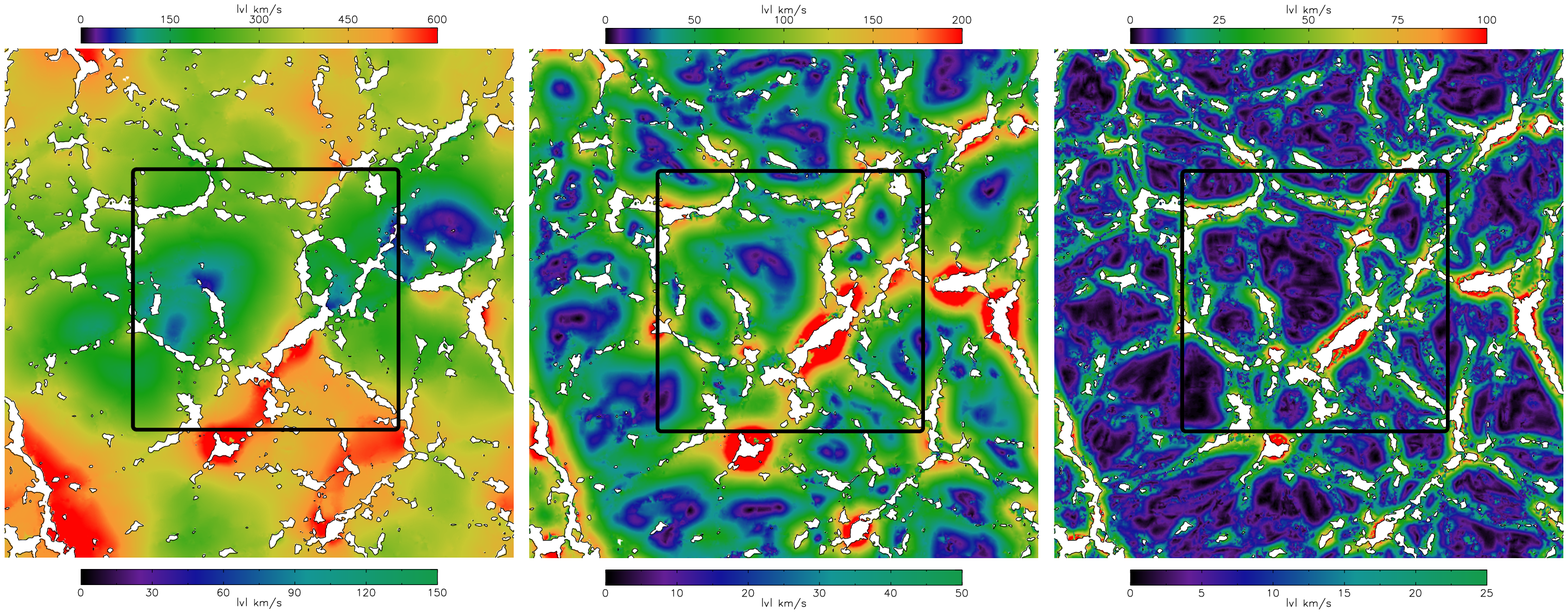}  
  \caption{Magnitude of the velocity field for the original velocity field (left panel), and the residual velocity field (see text for details) at scales 
                4 (center panel) and 1 $h^{-1}$Mpc (right panel). The magnitude corresponds to the full three-dimensional componets in
                contrast to Figure \ref{fig:hierarchical_velocities} where we only use the $x,y$ components.
  		The color bars on the top show the full dynamic range until the saturation point indicated by the highest value of
		the color bar. The lower color bar shows 1/4 of the full dynamical range. The matter distribution delineated by white filled isodensity contours
		at $\delta = 0$. The black square shows the region corresponding to 
		Figures \ref{fig:multires_density}, \ref{fig:multires_spine}, \ref{fig:hierarchical_velocities} \ref{fig:velocity_profiles} and
		\ref{fig:void_halo_vel}. Note tha the saturation point is 600, 200 and 100 km/s for the left, center and right panels respectively.}
\label{fig:velocity_magnitude}
\end{figure*} 
\section{Void dynamics}

We now focus on the internal dynamics of voids and their hierarchical character.  Figure \ref{fig:hierarchical_velocities} shows a slice of the
density field across the simulation box (top left panel) and the corresponding velocity field on the remaining panels. The velocity field was visualized using 
the \textit{particle advection} technique
in which test particles are randomly placed on the simulation box and their positions are evolved following
the instantaneous velocity field. The velocity fields shown here correspond to the two-dimensional velocity field
along the $x-y$ plane centered on the slice under consideration.  The structures in this two-dimensional field
are easier to visualize than in the three-dimensional case where one has structures along the line of sight which 
can lead to confusion. The advection field was created by iteratively displacing the test particles according to:

\begin{equation}
	\mathbf{x}_i ^{j+1} = \mathbf{x}_i^{j} + \mathbf{v}(\mathbf{x}_i^{j})_i^{j}
\end{equation}

\noindent where the index $j$ indicates the step in the advection process,  $\mathbf{x}_i ^{j}$ is the position of the particle $i$ and 
$\mathbf{v}(\mathbf{x}_i^{j})_i^{j}$ the velocity field evaluated at the current particle's position. Each particle is also assigned
a color based on their advection step $j$ according to a given lookup table.

From the slice in Figure \ref{fig:hierarchical_velocities} we select a region (yellow square) with a large void and focus on it. The corresponding 
unfiltered velocity field is shown in the right top panel. The structures in the velocity field have some correspondence to the
structures see in the density field. There is a weak correlation between the centers in the velocity field and the center of the large voids
although the boundaries of the voids are not delineated by the velocity field. The voids themselves seem to be embedded in a much larger 
bulk flow going from  the top right corner to the bottom-left corner  of the box. This is a consequence of the velocity field being dominated by 
fluctuations at larger scales. The corresponding amplitude of the velocity field is shown in Figure \ref{fig:velocity_magnitude}
(left panel). These large scale streams have amplitudes up to 600 km/s near massive structures. The bulk flow 
inside the central void of Figure \ref{fig:hierarchical_velocities} has velocities in excess of 300 km/s. There is a weak
correlation between the magnitude of the velocity field and the underlying distribution of matter.

\begin{figure*}
  \centering
  \includegraphics[width=0.99\textwidth,angle=0.0]{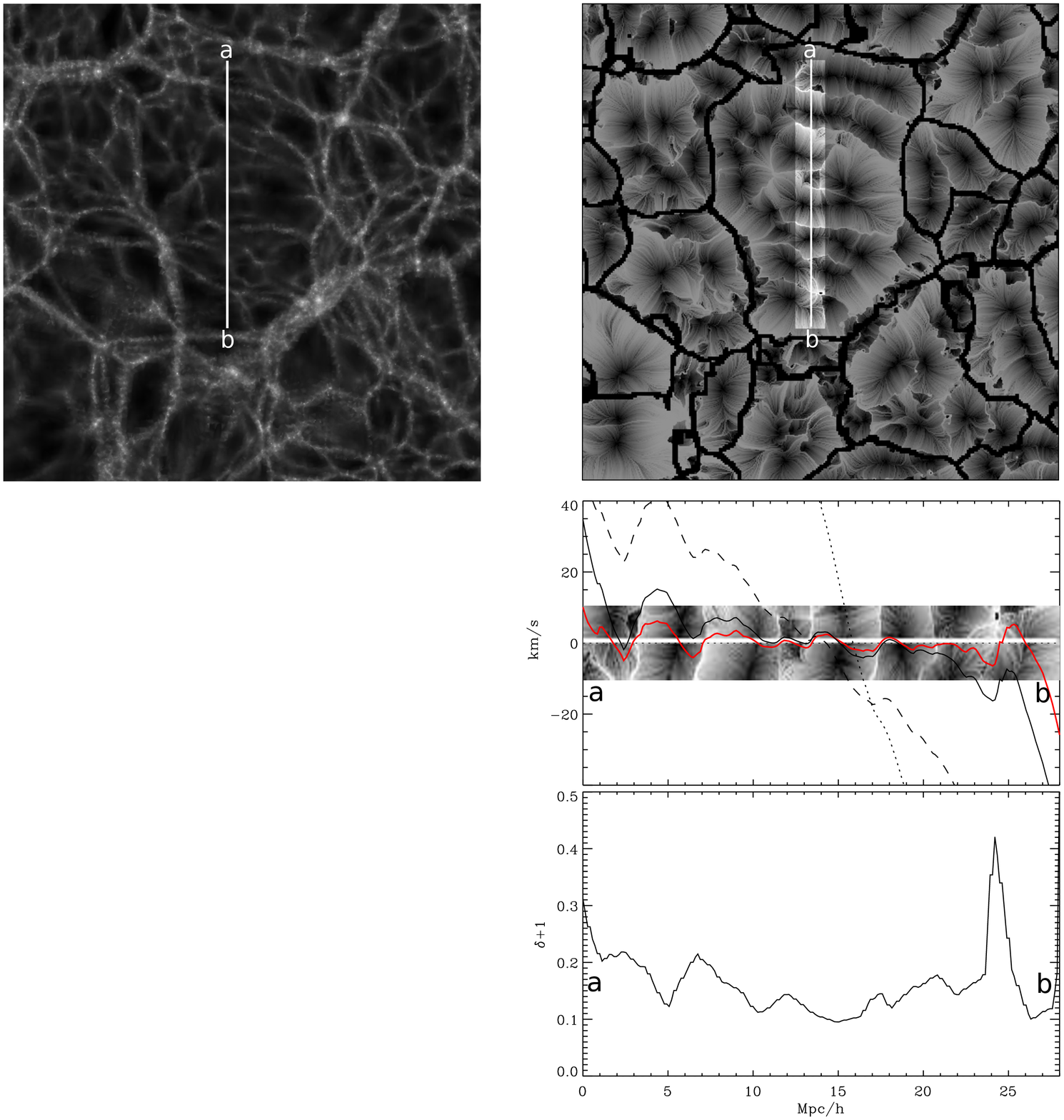} 
  \caption{Density and velocity profiles in the interior of a void. Top panels show a slice of the density field across
  the simulation box and the structures in the residual velocity field structures at scale 1 $h^{-1}$Mpc visualized with the particle advection
  technique (top left and right panels respectively).
  The top level of the hierarchical spine is superimposed on the residual velocity field.
  A vertical sampling line crosses the central void from point \textbf{a} to \textbf{b}. The sampled magnitude of the residual velocity field
  is shown in the
  middle right panel where the magnitude of the velocity field is decomposed into its residual components at scales $8,4,2$ and 1 $h^{-1}$Mpc
  (dotted dashed, solid black and solid red respectively). A small rectangle surrounding the vertical sampling line in the
  residual velocity field is highlighted in the top right panel and also shown in the middle right panel as the background for
  the residual velocity profiles in order to show the correspondence between the 1D profile at 1 $h^{-1}$Mpc (solid red line)
  and the 2D structures in the velocity field at the same scales. The bottom left panel shows the density field along the vertical 
  sampling line.}
\label{fig:velocity_profiles}
\end{figure*}

\subsection{Voids as hierarchically expanding domains}

In order to highlight the contribution to the velocity field from density fluctuations at scales relevant to voids we apply a high pass
filter to the velocity field (see Section \ref{sec:residual_velocities} for details). We choose filtering scales that fall below 
the peak in the void size distribution at $r < 10$ $h^{-1}$ Mpc (see Figure \ref{fig:void_sizes}) in order to probe the internal 
dynamics of voids. Figure \ref{fig:hierarchical_velocities} shows the residual velocity field at scales 4 and 1 $h^{-1}$ Mpc 
(right middle and bottom panels respectively). The structures in the residual velocity field now show 
a striking correspondence to the structures in the density field. The left middle panel shows the central void 
as a locally expanding domain with well defined boundaries closely matching their counterpart in the density field. 
Figure \ref{fig:hierarchical_velocities} shows voids more clearly identified in the residual velocity 
field than in the density field! This is a dramatic illustration of the shaping of the cosmic web 
and the dominant role of voids in the dynamics at these particular scales. 
The velocity field at scales of 4 $h^{-1}$Mpc is highly coherent and characterized by laminar flows. 
Only at the boundaries of voids the velocity field becomes turbulent
as we enter the dense regions dominated by non-linear interactions. 

Even more striking is the emergence of a hierarchy of structures in the velocity field as we probe smaller scales
(right bottom panel of Figure \ref{fig:hierarchical_velocities}).
With the residual velocity field at scale of 1 $h^{-1}$ Mpc the large central void seen as a coherent expanding 
domain at scale of 4 $h^{-1}$ Mpc fragments into several smaller voids. The nesting character of
the velocity field is (as fas as the authors know) shown here in an explicit way for the first time. 
The expanding sub-domains embedded in the larger ones correspond to sub-voids in the density field. 
By using the residual velocity instead of the original velocities we are able to study the contribution of different scales in the
cosmic web corresponding to different levels in the hierarchy of cosmic structures. Note that the parent void is perfectly delineated
by its sub-voids except  close to the boundaries where non-linear interactions make the velocity field turbulent.
In the present work we provide a qualitative description of the laminar or turbulent character of the velocity field. 
This will be addressed quantitatively and in more detail in a future paper.

The velocity field at the boundaries of the large voids tends to be highly turbulent reflecting the non-linear
nature of the dense filaments and clusters surrounding the voids. This can be better seen in the residual 
velocity field at scale 1 $h^{-1}$Mpc
(Figure \ref{fig:hierarchical_velocities}, bottom right panel). The level of turbulence in the
velocity field can be almost predicted by the local density (Figure \ref{fig:hierarchical_velocities}, top left panel).
This is not surprising since the high-density regions are dominated by small-range non-linear interactions. What is
more interesting is the velocity field at the boundaries of the structures delineated by the sub and sub-sub-voids.
The velocity field at the interior of the voids is practically laminar everywhere with the only exception of the
boundaries delineating the voids at the top of the hierarchy. The highly coherent velocity field inside voids
even at small scales reflects the fact  that the internal substructure of voids is still close to the linear regime and there has been
little shell crossing at the boundaries of the sub and sub-sub-voids.

The magnitude of the residual velocity field inside voids at scales of $\sim$ 1 $h^{-1}$Mpc (Figure \ref{fig:velocity_magnitude} right panel) 
is of the order of 10-20 km/s while at the interior of filaments at the same scales one can easily find velocities of $>$ 100 km/s.
This is a fundamental difference between voids and other environments. Inside voids at small scales the local dynamics are still
linear and strongly correlated to the underlying density field even at small scales. This may have important repercussions to the
mass accretion in void galaxies as we will see in coming sections.

\subsection{Velocity field and the Cosmic Spine}
 
The expanding domains in Figure \ref{fig:hierarchical_velocities} have some interesting properties. They form a contiguous network
of well defined, self-contained, locally expanding domains. There are no isolated structures in the velocity field just as in the density
field i.e. one does not find a disjoint filament inside a void neither in the velocity or in the density field. Even if a
structure appears not to be connected to anything, a higher-resolution inspection will always reveal it to be connected
by a tenuous web.
There is also no ``leaking" of matter between adjacent voids. Once matter reaches the void boundary it stays there as it continues 
its evolution towards walls, filaments and finally clusters. The picture that emerges from the velocity patterns in Figure 
\ref{fig:hierarchical_velocities} is that of a network of contiguous and closed cells.
The cellular nature of the Cosmic Web is perhaps the main reason behind the intimate relation between the geometry
and topology of the density field and why topological methods such as the Watershed Void Finder and the Cosmic Spine are able to 
identify geometrical structures in the density field by their topology.
The collapse of structures described by Zel'dovich generates a range of geometries but it is their context within the
Cosmic Web that defines their connectivity and topology as hinted by \citet{Bond96}. In Sections \ref{sec:hierarchical_emergence} 
and \ref{sec:hierarchical_spine} we discussed the motivation behind the
use of a hierarchical approach to the identification of structures. There we argued that the over segmentation problem in
the watershed void finder was not an intrinsic problem but (in our particular cosmological context) 
a direct consequence of the hierarchical nature of the cosmic web. It is then natural to expect that
every feature in the density field, provided the density field was properly reconstructed, should be identified to a real cosmic structure.
Figure \ref{fig:hierarchical_velocities} shows this very clearly. The top level of the Cosmic Spine hierarchy is shown 
on top of the residual velocity field and 4 $h^{-1}$Mpc (middle right panel). There is an almost perfect match between the
structures in the velocity field (and density field, see Figure \ref{fig:multires_density}). One must keep in mind that the
Cosmic Spine method makes no assumption on the geometry of the density field and only relies on the topology of the 
local minima. Its success in the identification of structure relies entirely on the close relation between the geometry
and topology of the cosmic web. If the cosmic web had different topological properties it would be impossible for the
Cosmic Spine to identify any feature, for instance any disjoint structure (e.g. an isolated filament/wall) would not
produce any change in the topology of the density field and would not be detected.

A careful inspection of the middle right panel of Figure \ref{fig:hierarchical_velocities} shows that there are
some voids that are not identified by the Cosmic Spine. These are structures that correspond to a different range in
scales than the one covered by this particular hierarchical level. The bottom panel of Figure \ref{fig:hierarchical_velocities}
shows the residual velocity field at scale 1 $h^{-1}$Mpc and the top and bottom levels of the Cosmic Spine.
Now the sub-domains are better correlated with the bottom level because we are in a different range of scales.
Even at this scales the top level Cosmic Spine is closely delineated by the much smaller
sub-voids.  

\subsection{Residual velocity field profiles}

A more direct comparison between the density and velocity fields is presented in Figure \ref{fig:velocity_profiles} where we show the residual
velocity profile at scales $8,4,2$ and 1 $h^{-1}$Mpc. As noted before the structures in the velocity field seem to
trace the hierarchy of voids more clearly than the density field by explicitly showing their expanding character.
Note the large variation between the different scales in the residual velocity field. The large scales have larger
amplitude as expected from the velocity power spectrum. 

The selected void has a density of $\delta \sim -0.9$ at its center and its density remains relatively constant until the
regions close to its boundaries. The velocity field on the other hand shows a strong variation as we cross the void with a clear
outflow pattern. This behaviour is seen only at scales $>$ 4 $h^{-1}$Mpc (dotted, dashed and black solid lines in middle panel of
Figure \ref{fig:velocity_profiles}) At small scales the residual velocity field has a similar behavior than the density field.

The residual velocity profile gives more information than a simple velocity profile where all the small-scale low-magnitude
details would be lost in the high magnitude of the raw velocity field. Note the correspondence in the magnitude of
the velocity profile at small scales (solid red line) and the 2D velocity structures in the top right panel. The solid red line 
crosses the $y$ axis at the origin and boundaries of the expanding domains in the background image. 
The same structures can be also seen in the density field (bottom right panel).

\begin{figure*}
  \centering
  \includegraphics[width=0.49\textwidth,angle=0.0]{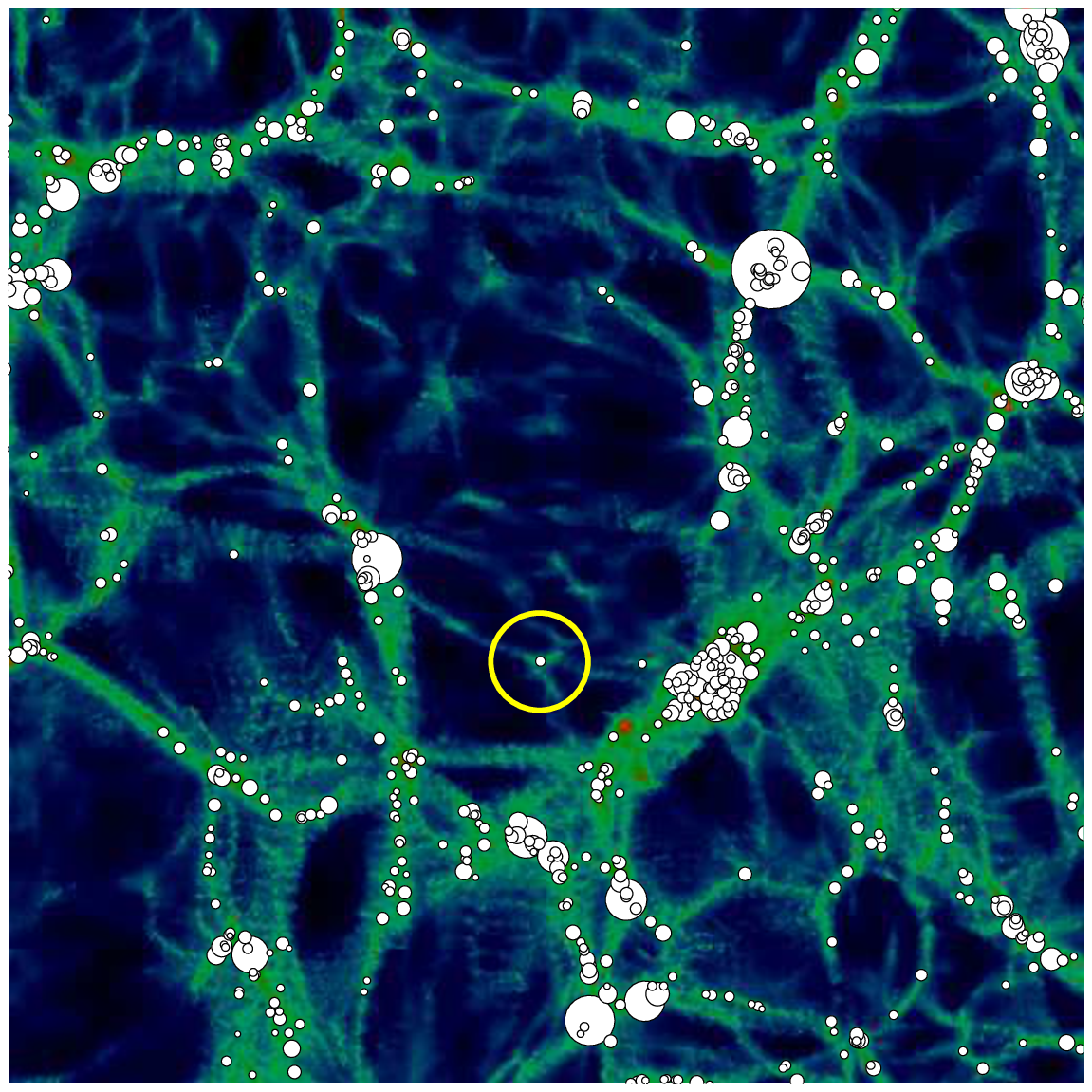}  
  \includegraphics[width=0.49\textwidth,angle=0.0]{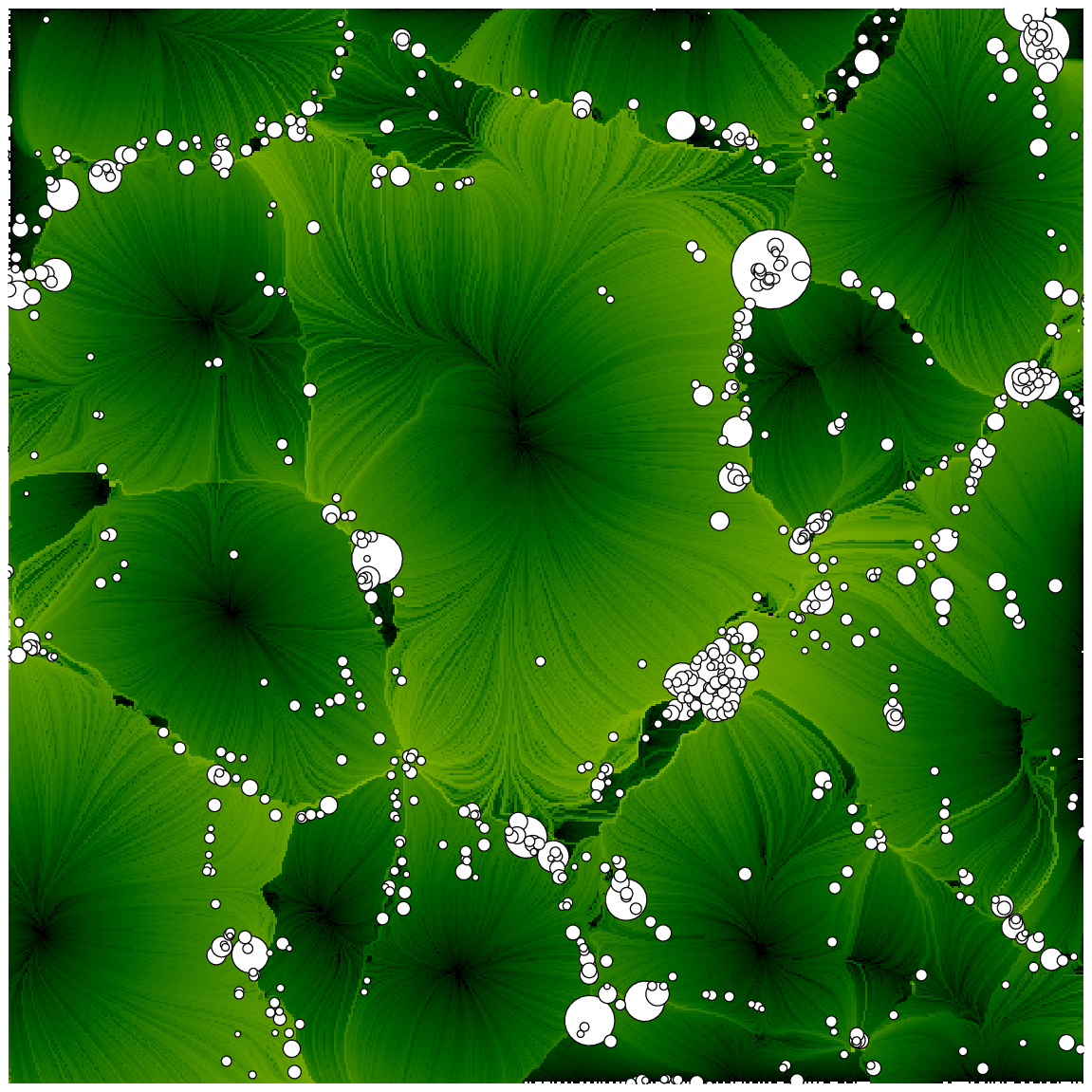}  
  \includegraphics[width=0.49\textwidth,angle=0.0]{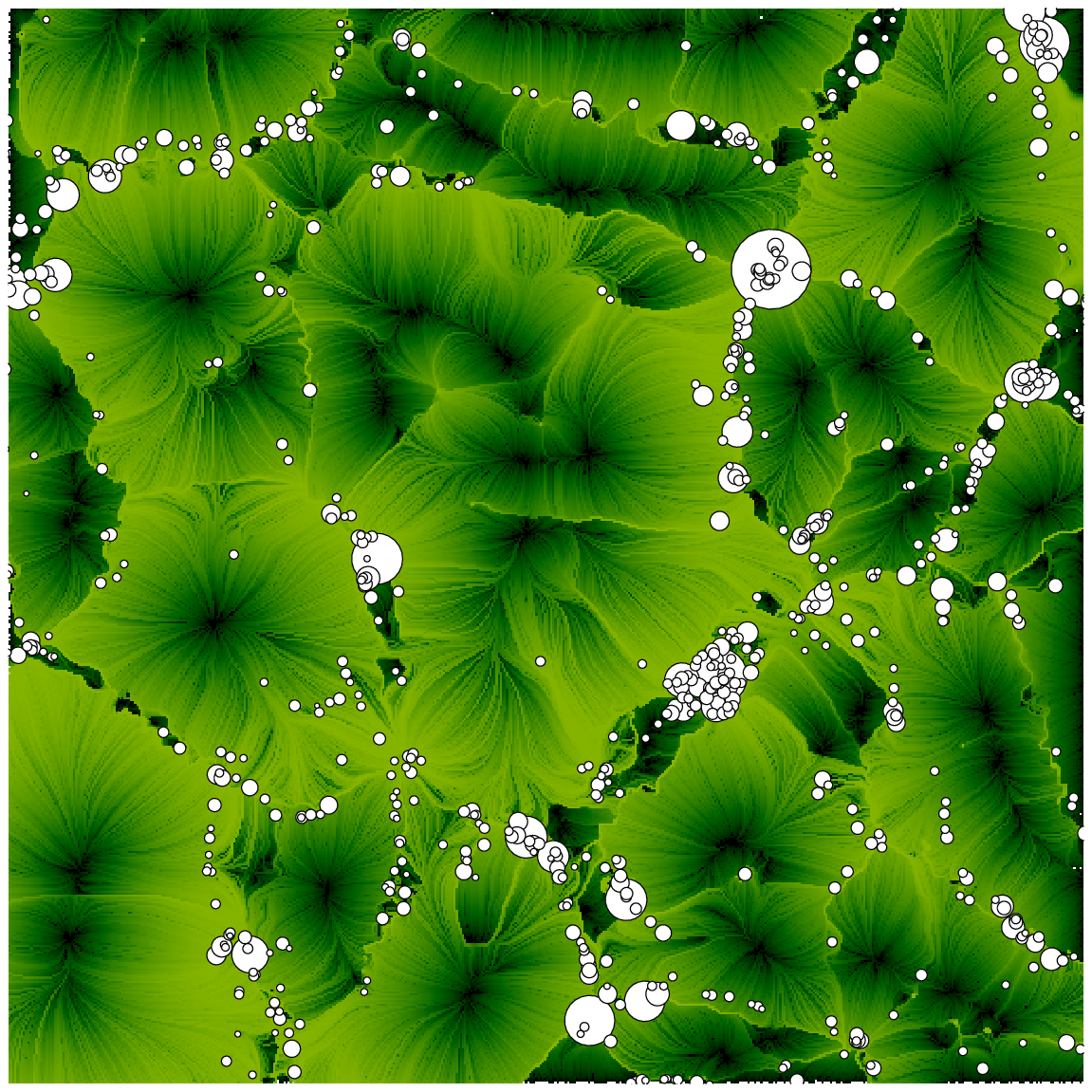}  
  \includegraphics[width=0.49\textwidth,angle=0.0]{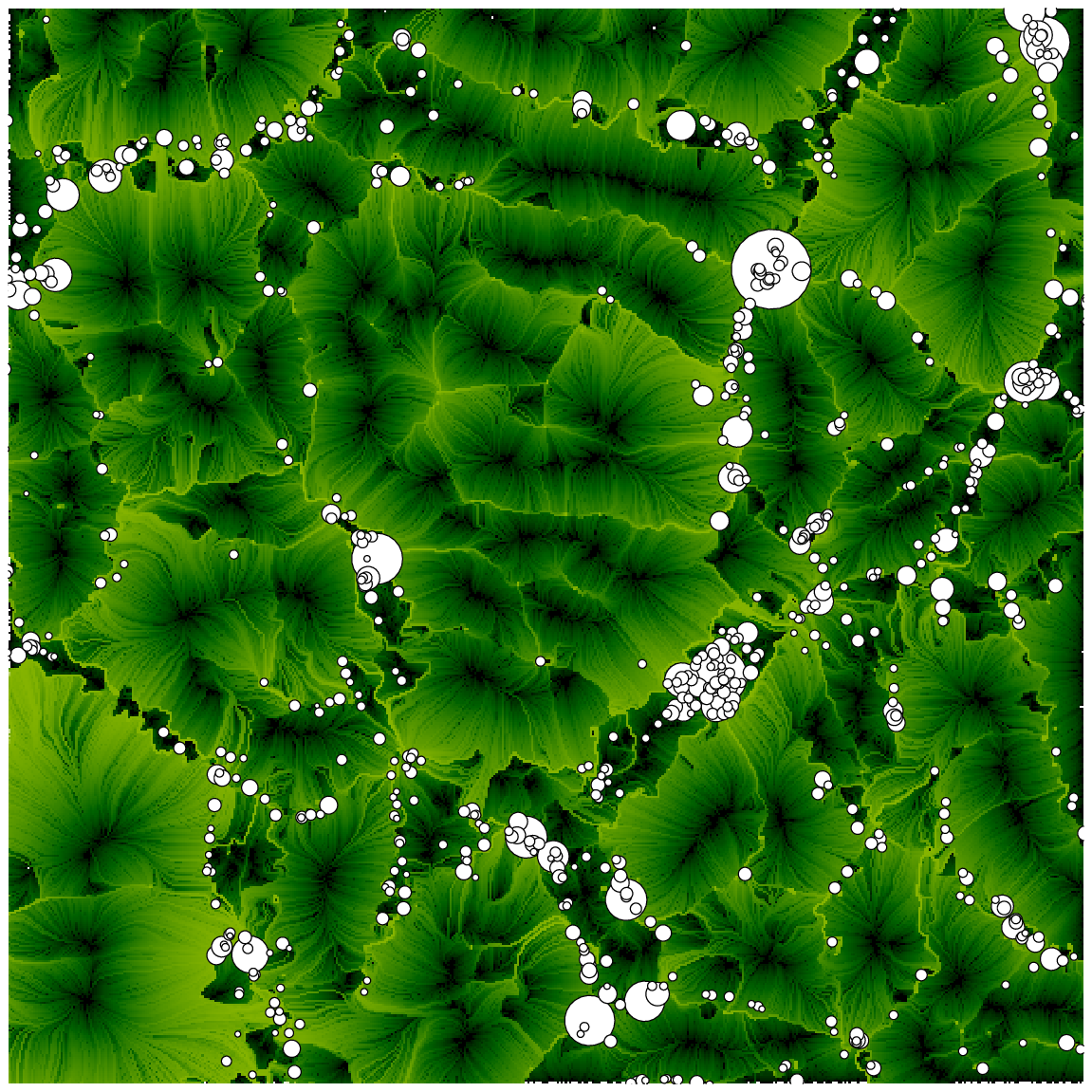}
  \caption{Dark matter haloes superimposed on the density and velocity fields. The let top panel shows a slice of the density field
      and on top of it all the FoF haloes closer than 1 $h^{-1}$Mpc from the slice.
      We select a halo inside the central void with a yellow circle (top left panel) for subsequent analysis (see text for details)
	  The residual velocity field is shown at scales 4,2 and 1 $h^{-1}$Mpc in the top right, bottom left and bottom right respectively.
	  We show the same FoF haloes as in the density field slice. The velocity field was visualized using the particle advection technique
	  (see text for details).}
	\label{fig:void_halo_vel}
\end{figure*}

\section{Void galaxies}

In the previous section we have seen that the velocity field inside voids is a superposition of hierarchically 
expanding domains dominated by large scales flows. Now we focus on the small scale velocity field that defines the
local dynamics around the low-mass haloes found inside voids. 

Figure \ref{fig:void_halo_vel} shows a slice of the density field (top left panel) and the FoF haloes identified 
inside a slice of 2 $h^{-1}$Mpc of thickness. The haloes are arbitrarily scaled with their virial radius.
Most of the haloes in the slice are located in filaments and clusters. There is a strong
correlation between the mass/size of the halo and its local density. The 
most massive haloes invariable occupy the dense regions and completely avoid the interior of voids. Instead voids are 
populated by low-mass haloes as one would expect given their underdense environment.
This low-mass population is the combined result of the low-density environment which
gives a slower gravitational development and the super-Hubble local dynamics around the
haloes that decreases the radius from which the halo can accrete mass \citep{Aragon11a}.

In what follows we focus on the central void in Figure \ref{fig:void_halo_vel} from which we select a halo far from its boundaries 
(the halo is shown in the top left panel enclosed in a yellow circle). The void halo has a 
mass of $2.79 \times 10^{10}$ $h^{-1}$ M$_{\odot}$ and a radius of 89 $h^{-1}$kpc. 
Figure \ref{fig:void_den_profile} shows the density inside increasing spheres center in the void halo. At very small radius the 
enclosed density profile is dominated by the compact halo but at $r \sim$ 1 $h^{-1}$Mpc it rapidly falls to the mean
void density at $\delta \sim -0.9$. After this point the profile slowly increases as the denser nearby walls and
filaments are included in the sampling volume. This profile does not however shows the ``real" profile of the void because 
the halo is not located at its center. At a distance from the halo of 8 $h^{-1}$Mpc the enclosed density reaches again the mean 
density and continues increasing until it reaches its peak at R$ \sim 9$ $h^{-1}$Mpc. If we ignore the
contribution of the halo at small radius this would be a textbook void profile.
This halo is a classical example of the \textit{cloud-in-void} scenario. Compare Figure \ref{fig:void_den_profile} with the second panel in 
Figure 6 of  \citet{Sheth04}. Both curves have a similar behavior even though our profile was computed from
the present-time density field.

\subsection{Matter accretion in void haloes}

The dynamical environment around haloes is shown starting at the top right panel in Figure \ref{fig:void_halo_vel} and
continuing clockwise for the residual velocity field at scales 4,2 and 1 $h^{-1}$Mpc respectively. At scales 4 and 2 $h^{-1}$Mpc the velocity field
around the void halo is basically laminar and the halo is simply dragged away from the void in the bulk flow. 
These flows have magnitudes of the order of 50-100 km/s (see Figure \ref{fig:velocity_magnitude}).

At smaller scales the velocity field sudlendly presents a dramatic change. Instead of a large-scale laminar flow we now
have small-scale coherent streams emanating from the centers of very small adjacent voids and reaching the
halo in an anisotropic pattern. The small voids around the halo define a web of tenuous filaments
with the halo sitting at their intersection. This structure can be seen in both the density and velocity field
but the velocity field explicitly shows how the matter is locally shaped into this anisotropic configuration.
This low mass halo is the dominant component in the local dynamics in a similar way as a cluster with a mass several orders 
of magnitude larger would be at much larger scales. The halo also accretes mass from its connected tenuous filaments again resembling the
accretion of matter in massive clusters.
Inside the void clean streams of matter feed the halo at very small rates. the residual velocity field at 1 $h^{-1}$Mpc 
has a magnitude of the order of 10-20 km/s in contrast to the dense megaparsec-long
filaments where it can be as high as 100 km/s at same scales (See Figure \ref{fig:velocity_magnitude}).

The most striking feature of the velocity field around void galaxies is its highly anisotropic and coherent nature.
The local mini web around void haloes marks a fundamental difference with similar haloes in denser environments. Here the halo
dominates the local dynamics allowing it to feed from a well defined tenuous web. A similar halo located inside a filament will
be embedded inside the larger dense filament with non-linear turbulent dynamics. The high coherence and low magnitude of the
velocity field around void halos may help explain the sustaining of the delicate extended gas disks characteristic of ``isolated" 
galaxies \citep{Kreckel11a} and the also tenuous ``rings" of gas observed around galaxies in low-density environments
\citep{Stanonik09}.

\begin{figure}
  \centering
  \includegraphics[width=0.49\textwidth,angle=0.0]{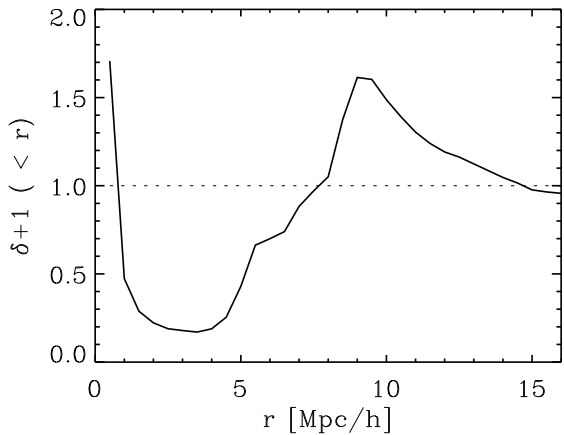}
  \caption{Enclosed density profile centered in the halo selected in Figure \ref{fig:void_halo_vel}.}
  \label{fig:void_den_profile}
\end{figure}

While the particle resolution of our simulation does not allows us to make accurate measurements of mass accretion we
can make a first order estimation for this particular halo based on the local density and velocity field. The local web 
around the halo drags matter from its surroundings effectively increasing the local density.
From Figure \ref{fig:void_den_profile} we can safely assume a local density of 0.5 the mean density of the Universe 
in the immediate neighborhood of the halo.
The velocity field around the halo is of the order of $\sim 20$ km/s which gives a gas accretion rate inside a
sphere of 1 $h^{-1}$Mpc centered on the halo of $\dot{m}_{\textrm{\tiny{gas}}} \simeq 1.25$ M$_{\odot}$ yr$^{-1}$ assuming a barion 
fraction of 0.04. If we assume that most of the accreted gas is converted into stars then we have similar values for star formation rates than 
those obtained in high-resolution simulated void haloes \citep{Kreckel11d}. Perhaps more important than the magnitude of the
accretion rate is the steady and coherent nature of the accretion process which provides a highly efficient mechanism
to inject cold gas into the inner region of the halo \citep{Keres05,Dekel09}.


\section{Conclusions and discussion}\label{sec:conclusions}

In this paper we introduced the most recent implementation of the Hierarchical Cosmic Spine method  and applied it to a cosmological N-body 
simulation in order to decompose the density field into its basic geometrical components. We focused our analysis on voids identified
on a three-level hierarchical space. 

The size distribution of voids has a log-normal behavior in agreement with previous findings \citep{Platen07}. Here we went one step further
and computed the void size distribution for the three hierarchical levels computed in this work. At the top of the hierarchy 
the peak of the void size distribution is $\sim 8$ $h^{-1}$Mpc which is very similar to estimates of the radius of the local void \citep{Nasonova11}
indicating that the local void is average in size. The size distribution of sub-voids and sub-sub-voids have peaks at smaller scales as
expected from their hierarchical nature. The void size distribution suggest that these structures are delineated by galaxies with luminosities
in the range -16 $<$ M$_r$ $<$ -20 according to \citep{Furlanetto06}. 

Voids are structures with very complex internal dynamics. By removing the large-scale component of the velocity field we 
illustrate how the hierarchy of voids in the matter distribution is reflected in a hierarchy of expanding domains in the velocity field. 
The notion of voids as simple expanding domains is far from accurate. Instead, voids have a rich internal hierarchy
of expanding sub-domains where the main expansion defining the parent void breaks down into smaller expanding sub-domains 
at smaller scales and so on. Furthermore we also showed that the structures seen in the density and
velocity field as well as their hierarchy are successfully identified by the Hierarchical Cosmic Spine method.
Our method is unique in that it not only identifies structures in a hierarchical way but it also gives their explicit hierarchical 
relations.
			
The velocity field inside voids can be characterized as a laminar flow dominated by large scales. The fluctuations in the
residual velocity field at scales $<$ 1 $h^{-1}$ Mpc are one order of magnitude lower than the fluctuations at scales comparable to the
void size. The overall internal dynamics of voids are largely dominated by scales $> 4$ $h^{-1}$Mpc while the smaller scales play a minor role. 

We presented an exploratory study of haloes inside voids. Contrary to voids haloes are not affected by the large scale motions. They
simply move together with their environment in large-scale and large-amplitude bulk flows. Small scales on the other hand play a major role
by defining the dynamics around the halo and the properties of the galaxies they host. We found that haloes inside voids feed through 
highly coherent laminar streams of matter in sharp contrast with the environment inside filaments and clusters where
the velocity field is turbulent. These streams produce a highly anisotropic matter accretion pattern around the
halo and may provide an efficient gas feeding mechanism with low accretion rates that can be sustained for large periods of time.
This steady and coherent gas accretion may offer a natural explanation for some of the observed properties of galaxies in voids and
low-density environments such as extended gas disks, polar disks, higher gas content and star formation rates. I a forthcoming paper 
we will explore in more detail the effect of the small-scale velocity field in the gas accretion of galaxies in voids and how 
this affects their observed properties.

Voids represent a unique opportunity to probe the structure and dynamics of the early Universe at small scales by doing local observations.
Future ultra-deep surveys covering the volume of nearby voids may one day unveil their rich substructure and dynamics. In this direction
the Void Galaxy Survey \citep{Kreckel11b} is leading the way in the study of this fascinating objects with already some tantalizing
results like the possible finding of the first inter-void mini filament in the Ursa Minor I void seen as a long tenuous bridge of matter 
connecting three galaxies (VGS-38 in Figure 7 of \citet{Kreckel11b}).

\section{Acknowledgements}

Miguel A. Aragon-Calvo would like to thank Mark Neyrinck for many stimulating discussions. This work was funded by the
Gordon and Betty Moore Foundation.

\end{document}